\documentclass[a4paper,11pt]{article}
\pdfoutput=1 

\usepackage{jcappub} 
\usepackage[utf8]{inputenc}
\usepackage[T1]{fontenc}
\usepackage{xcolor}

\usepackage{bm}
\let\vec\bm
\let\ve\bm
\usepackage{soul}

\newcommand{\Dk}[1]{\frac{d^3#1}{(2\pi)^3}}

\newcommand{\vk}{\ve k}
\newcommand{\vp}{\ve p}
\newcommand{\vq}{\ve q}
\newcommand{\vx}{\ve x}

\newcommand{\vv}{\ve v}
\newcommand{\vs}{\ve s}
\newcommand{\vu}{\ve u}
\newcommand{\vr}{\ve r}

\newcommand{\dD}{\delta_\text{D}}
\newcommand{\Ps}{\mathbf{\Psi}}

\newcommand{\vhn}{\hat{\ve n}}

\def\apjl{ApJL}
\def\apjs{ApJS}

\def\aap{{A\&A}}

\usepackage{booktabs}
\usepackage{float} 
\newcommand{\ra}[1]{\renewcommand{\arraystretch}{#1}}

\title{\boldmath Full Shape Cosmology Analysis from BOSS in configuration space using Neural Network Acceleration}

\author[a]{Sadi Ramirez}
\author[b, {\color{blue} \dagger}]{Miguel Icaza-Lizaola\note{ {\color{blue} $\dagger$ Corresponding author.}}}
\author[c]{Sebastien Fromenteau}

\author[a]{Mariana Vargas-Magaña}
\author[c, d, e]{Alejandro Aviles}

\affiliation[a]{Instituto de Física, Universidad Nacional Autónoma de México, Apdo. Postal 20-364, 01000, D.F, México.}
\affiliation[b]{Korea Astronomy and Space Science Institute,  776 Daedeok-daero,
Yuseong-gu, Daejeon 34055, Republic of Korea.}
\affiliation[c]{Instituto de Ciencias F\'isicas, Universidad Nacional
Autónoma de México,  62210, Cuernavaca, Morelos.}

\affiliation[d]{Departamento de F\'isica, Instituto Nacional de Investigaciones Nucleares,
Apartado Postal 18-1027, Col. Escand\'on, Ciudad de M\'exico,11801, M\'exico.}

\affiliation[e]{Consejo Nacional de Ciencia y Tecnolog\'ia, Av. Insurgentes Sur 1582,
Colonia Cr\'edito Constructor, Del. Benito Ju\'arez, 03940, Ciudad de M\'exico, M\'exico.}


\emailAdd{sadiramirez@estudiantes.fisica.unam.mx}
\emailAdd{icaza@kasi.re.kr}
\emailAdd{sfroment@icf.unam.mx}
\emailAdd{mmaganav@fisica.unam.mx}
\emailAdd{avilescervantes@gmail.com}

\abstract{
Recently, a new wave of full modeling analyses have emerged within the Large-Scale Structure community, leading mostly to tighter constraints on the estimation of cosmological parameters, when compared with standard approaches used over the last decade by collaboration analyses of stage III experiments. However, the majority of these \textit{full-shape} analyses have primarily been conducted in Fourier space, with limited emphasis on exploring the configuration space. Investigating n-point correlations in configuration space demands a higher computational cost compared to Fourier space because it typically requires an additional integration step. This can pose a limitation when using these approaches, especially when considering higher-order statistics. One avenue to mitigate the high computation time is to take advantage of neural network acceleration techniques. In this work, we present a  full shape analysis of Sloan Digital Sky Survey III/BOSS in configuration space using a neural network accelerator. We show that the efficacy of the pipeline is enhanced by a time factor $10^{3}$ without sacrificing precision, making it possible to reduce the error associated with the surrogate modeling to below $10^{-2}$ percent which is compatible with the precision required for current stage IV experiments such as DESI.  We find   $\Omega_m=0.286\pm 0.009$,   $H_0=68.8\pm 1.2$ $\mathrm{km} \mathrm{s^{-1}}\mathrm{Mpc^{-1}}$ and $A_s \times 10^9  =2.09 ^{+0.25}_{-0.29}$. Our results  on public BOSS data are in good agreement with BOSS official results and compatible with other independent full modeling analyses. We explore relaxing the prior on $\omega_b$ and varying $n_s$, without significant changes in the mean values of the cosmological parameters posterior distributions, but enlarging their widths. Finally, we explore the information content of the  multipoles when constraining cosmological parameters.

}

\begin{document}
\maketitle
\flushbottom

\section{Introduction}
\label{sec:intro}
Over the last decade, the study of the clustering of galaxies through Large Scale Structure (LSS) surveys  has emerged as a crucial probe within precision Cosmology. Spectroscopic surveys, such as the Sloan Digital Sky Survey\footnote{\href{https://www.sdss.org/}{www.sdss.org}} (SDSS) and  the Dark Energy Spectroscopic Instrument\footnote{\href{https://www.desi.lbl.gov/}{www.desi.lbl.gov}} (DESI) provide three-dimensional maps of the Universe, where angular positions and redshifts of millions of galaxies are measured with high accuracy. These maps constitute appropriate data sets for quantifying the clustering characteristics of galaxies, including correlation functions and other summary statistics, which then can be compared to models' theoretical predictions. 

The study of clustering statistics has primarily relied on two significant sources of cosmological information. First, Baryon Acoustic Oscillations (BAO) in the early Universe freeze up at the drag epoch, whose signature can be observed at later times in the correlation function as a well-distinguished peak around a scale of 150 $\textrm{Mpc}$. Second, peculiar velocities of galaxies contributes to the redshift that we measure adding to the Hubble flow component. Since this contribution occurs only along the line-of-sight direction, we observe an apparent anisotropic distortion in the matter distribution, and hence, galaxy statistics which otherwise would be isotropic become dependent on the angle of observation. This effect is known as Redshift Space Distortions (RSD).

RSD and BAO effects most of the relevant information about the correlation function of galaxies. Consequently, the SDSS-III BOSS \citep{2011AJ....142...72E,2013AJ....145...10D,2015ApJS..219...12A} and SDSS-IV eBOSS \citep{2021MNRAS.500.3254R,2020MNRAS.498.2354R,2020ApJS..250....8L} collaborations have chosen compressed methodologies for their standard analysis. In such approaches, the cosmological parameters of the matter power spectrum are fixed to fiducial values, and a set of parameters characterizing the BAO and RSD effects are explored. Given that RSD depends on the average velocity of galaxies, it is sensitive to the growth rate of structure, so one of the chosen parameters is $f\sigma_8$. Two additional degrees of freedom should be included to account for the distortions in the position of the BAO along and across the line-of-sight, which arise from potential mismatches between the fiducial and true cosmologies, that is,  the Alcock-Paczy\'nski effect \citep{1979Natur.281..358A}.

On the other hand, the Effective Field Theory of LSS (hereafter simply EFT) \cite{McDonald:2006mx,McDonald:2009dh,Baumann:2010tm,Carrasco:2012cv,Vlah:2015sea,Chen:2020zjt} built on top of Perturbation Theory \cite{Bernardeau:2001qr} has been developed during the past years, and by now is currently used in analyses that confront theoretical models of the galaxy distribution directly to the data gathered by our telescopes. These methods are commonly known as full-modeling or full-shape analyses, and operate in a similar fashion that it has been done for the CMB over the years.  Nowadays, these full-shape templates are used routinely to constrain cosmological parameters \cite{Ivanov:2019pdj,DAmico:2019fhj,Philcox_2020,Chen:2021wdi, Zhang2022,
Donald-McCann:2023kpx}, including higher-order statistics \cite{Philcox:2021kcw,Philcox:2022frc} and even beyond $\Lambda$CDM models \cite{Chudaykin:2020ghx,DAmico:2020tty,Carrilho:2022mon,Piga:2022mge,He:2023oke}. 

One of the primary advantages of the compressed methodology is its agnostic nature: the parameters it explores are relatively model-independent when compared with those obtained from a direct, full-shape analysis. However, generating full-shape theoretical templates of the power spectrum comes with significant computational costs, which, until recently, hindered our ability to perform cosmological parameter estimation. This is one of the reasons why the compressed methodology has been favored by some part of the community. But even if the full-shape analysis is expensive and model-dependent, it is capable of extracting more cosmological information from the power spectrum.  Therefore, both the compressed and full-shape methodologies have their own merits, and there should be an incentive to pursue both approaches in parallel.

As we transition into a new era of cosmological surveys, the costs of full-shape models are set to increase even further. This is due to the unprecedented precision achieved in measurements on small scales of the correlation function (smaller than 50 $h^{-1}\,\textrm{Mpc}$), where the nonlinearities of perturbations exert a none negligible influence on halo distributions. Furthermore, at these small scales, the relationship between the clustering of observable astrophysical tracers and the underlying dark matter halos becomes complex. As a consequence, building accurate templates at these small scales might require the evaluation of even more complicated models, thereby introducing more intricate calculations and increasing the computational cost of an individual template. Finally, the analytical modeling of higher order statistics will pose new computational time challenges.  

In recent years, a number of avenues have been developed to tackle these difficulties and going beyond the standard three parameters compress methodology and into more intricate models at this smaller scales. One possible road is to expand the compress approach by introducing a small subset of new free parameters that encompass most of the remaining relevant information in the power spectrum \cite{2021JCAP...12..054B}. Another possibility is to perform full shape analyses encompassing various cosmological parameters. To achieve this, several optimizations in the computational methods used to construct theoretical templates have been developed, significantly reducing the computational cost of full shape analyses. 

As a result, many groups have reanalyzed the BOSS and eBOSS data using full-shape modeling. The optimized methodologies employed for these analyses can be categorized into two groups: efficient theoretical templates of the power spectrum \cite[e.g.][]{Ivanov:2019pdj,DAmico:2019fhj, Philcox_2020,Troster_2020,Chen_2022,Semenaite_2023} or emulator techniques that learn how to reproduce expensive models \cite[e.g.][]{Neveux_2022}. 
The consensus from all of these reanalysis methods is that the constraints on cosmological parameters like the Hubble constant $H_0$ are significantly tighter when using these improved approaches \cite{Maus_2023}. 

In configuration space, the number of analyses in the literature is reduced, mainly because the consensus is that direct-fits are more constrictive in Fourier space. Nevertheless,  working directly in configuration space has its benefits, particularly when dealing with the well localized BAO peak, which in Fourier space becomes distributed across a wide range wave numbers. Adding to that, some of the observational systematic have different effects in Fourier and configuration space. Consequently there is an incentive to study both spaces simultaneously. In ref.\cite{Chen_2022} the main analysis is performed in Fourier space but the authors have also worked out the correlation function as a consistency check. On the other hand, the work of \cite{Zhang2022} is devoted to perform fitting only to the correlation function using the \textsc{PyBird} code. The main difference with our approach is that we work from the beginning in a Lagrangian framework, and get directly the correlation function without the necessity of obtain first the power spectrum (plus adding infrared resummations) as an intermediate steps, to at the end Fourier Transform the results to the obtain the correlation function.

The number of free parameters to explore in full shape analysis is generally large when compared to the standard approach. With this in mind, considerable efforts have been made in building efficient sampling methods \cite{Foreman-Mackey:2012any, zeusensemblekaramanis2020, zeuskaramanis2021} which reduce the number of model evaluations required to run Markov Chain Monte Carlo (MCMC) explorations. However, the models will most likely continue to increase in complexity in the next years, since higher loop contribution or higher order statistics can be considered in the analyses. Therefore, there is an incentive to reduce the computational cost of generating these theoretical templates.

Machine Learning algorithms like Neural Network have been successfully used to drastically reduce the evaluation times of complex models \citep[e.g.][]{2022JCAP...04..056D}. These techniques use datasets of pre-computed templates at different points within the parameter space and learn how to reproduce them. When trained correctly, neural networks can reproduce fairly complex models with an error smaller than the precision needed by LSS surveys. Also, given that neural networks are not local interpolators, in principle the errors in their predictions are not as strongly dependent on the distance of the nearest point within the training set, as methodologies like Gaussian processes emulators would be. 

In this work, we model the redshift space correlation function up to one-loop perturbation theory using a Gaussian Streaming model \cite{Fisher:1994ks,Scoccimarro:2004tg,Reid:2011ar,Wang:2013hwa,Vlah:2015sea} in combination with Effective Field theory (EFT). Throughout this work, we refer to our modeling as EFT-GSM. To implement our model, we release a code\footnote{\href{https://github.com/alejandroaviles/gsm}{https://github.com/alejandroaviles/gsm}} that uses a brute force approach, but still can compute the correlation function in $\mathcal{O}(1 \, \text{sec})$ time, as described in section \ref{subsec:code}. However, the number of evaluations required to build a convergent MCMC chain for our baseline analysis is large, and this process can take a considerably amount of time. Therefore, we also built a neural network emulator to accelerate the computation of individual templates, reducing the running time of an MCMC chain from a few tens of hours to around 60 minutes (using the same computing settings) and below 20 minutes when we run the code in parallel depending on the cluster settings. 

We utilize our methodology to reanalyze the BOSS DR12 LRG data obtaining the tightest constraints on the $\Lambda$CDM parameters using the 2-point correlation function alone. Throughout this study, we pursue two primary objectives. The first is to bring forward the potential of our full shape modelling approach in configuration space for extracting cosmological information when compared to its counterpart in Fourier space. The second objective is to demonstrate that neural network surrogate models can be used safely to optimize cosmological analysis, leading to significant savings in both time and computational resources without sacrificing accuracy when analyzing real data.

We have also tested extended regions in parameter space compared with the baseline analysis to include scenarios with prior configurations beyond Planck \cite{Planck:2018vyg} and Big Bang Nucleosynthesis (BBN) \cite{Cooke_2018} on parameters $\Omega_b$ and $n_s$ that will serve us to explore the potential of LSS observables standalone. This serves as a test of the full methodology in highly more degenerated scenarios than the baseline and with aim to prove the viability of the use of neural network in this larger and more complex parameter space. 

This paper is organized as follows, we begin in section \ref{sec:data,sim}, introducing the data from the BOSS collaboration that we analyze here, as well as introducing a set of different mock simulations that we use for testing our methodology and building the covariance matrices required for our likelihood estimations. Then in section \ref{sec:model} we introduce the EFT-GSM model that we utilize to construct our theoretical templates.  Then, in section \ref{sec:NN_definition} we introduce our neural network methodology, which is used as a surrogate model instead of the EFT-GSM model, we also quantify how much efficiency is gained with this. In section \ref{sec:method} we describe the fitting methodology, including a brief description of full shape fits. Here we also emphasise our parameter space and the priors that we impose on each parameter. Section \ref{sec:testmocks} presents the validation of the methodology with high precision mocks and section \ref{sec:results} presents the results of our baseline analysis and how it compares with published alternative analysis of BOSS. We also include a subsection with results expanding the priors for cosmological parameters usually constrained independently by other observables and a subsection exploring the information content in the multipoles.

\section{Data and Simulations}
\label{sec:data,sim}
\subsection{Data}
\label{sec:data}

We analyse the publicly available data from Sloan Baryon Oscillation Spectroscopic Survey (BOSS) \cite{,2013AJ....145...10D}, which was a part of the  Sloan Digital Sky Survey III \cite[SDSS-III;][]{2011AJ....142...72E}. Specifically, we utilize the Data Release 12 galaxy catalogues \cite{2015ApJS..219...12A}, gathered using the 2.5-meter telescope situated at the Apache Point Observatory in New Mexico, USA \cite{2006AJ....131.2332G}, and all the spectra were measured using a set of multi-object spectrographs \cite{2013AJ....146...32S}. The details about the data reduction methodology can be found at \cite{2012AJ....144..144B}.

The BOSS target selection was designed to collect data for two different samples: the low-redshift sample (LOWZ), which targeted luminous red galaxies at redshifts $z<0.4$, and the Constant Stellar Mass sample, (CMASS), which targeted massive galaxies in the $0.4<z<0.7$ redshift range. As explained in \cite{2014MNRAS.441...24A,2017MNRAS.470.2617A}, LOWZ and CMASS samples were later combined into three partially overlapping bins, this was done to optimise obtaining the strongest constraints on the dark energy parameters. Throughout this work we will refer to these bins as $z_1$, $z_2$ and $z_3$ respectively. The catalogue construction is described in \cite{Reid2016,Ross2017,2017MNRAS.470.2617A} where the masks, completeness, and weights of the sample are also discussed. The main properties of these samples are summarized in Table \ref{table:data}. 

Our final analysis is performed using the low and high redshift  bins ($z_1$ and $z_3$, respectiveely) which do not overlap in redshift and have a similar effective volume, $V_\mathrm{eff}$.\footnote{The effective volume is defined by $V_{\mathrm{eff}}=\sum_i \left ( \frac{\bar{n}(z_i) P_0}{1+\bar{n}(z_i)P_0} \right )^2 \Delta V(z_i)$ where $\Delta V(z_i)$ is the volume of the shell at $z_i$ with $P_0=10,000 h^{-3}\mathrm{Mpc}^3$. The value of $P_0$ is chosen for being the amplitude of the power spectrum where the BAO signal is larger \cite{Reid2016, Anderson2014}.} 
\begin{center}
\begin{table*}
\ra{1.7}
\begin{center}
 \begin{tabular}{c c c c c c}

 \toprule
$\,\,$ Name $\,\,$  & $\,\,$ $z$-range  $\,\,$ & $\,\,$ $z_\mathrm{eff}$ $\,\,$ & $\,\,$  $N_\mathrm{gal}$  $\,\,$ & $\,\,$  $V_\mathrm{eff}$ $\,\,$ & $\,\,$  $V$  $\,\,$ \\ 
    \hline

 LOWZ   &  $0.15 < z < 0.43$   &   0.32   &   361,762  &  2.87 &  3.7 \\ 
CMASS   &   $0.43 < z < 0.70 $   &   0.57   &   777,202  &  7.14  & 10.8 \\ 
BOSS $z_1$  & $0.20 < z < 0.50$     & 0.38   &  604,001  &  3.7  & 6.4\\ 
BOSS $z_2$  & $0.40 < z < 0.60$     & 0.51   & 686,370   &  4.2 & 7.3 \\ 
BOSS $z_3$  & $0.50 < z < 0.75$     & 0.61   & 594,003   &  4.1 & 12.3 \\               
\hline
\end{tabular}
\caption{Summary of the properties of the samples used in this study. The effective volume
considers $P_0=10,000 h^{-3}\mathrm{Mpc}^3$,  and a fiducial cosmology with $\Omega_m=0.310$ and $h=0.676$. } 
\label{table:data}
\end{center}
\end{table*}
\end{center}

\subsection{Simulations}
\label{subsec:simulations}

In this work, we employ two distinct sets of simulations that we require for constructing the necessary covariance matrices for our likelihood estimations (see section \ref{sec:method}), and for validating our methodology using high-precision mocks. We now present a brief overview of these simulations. 

\begin{itemize}
\item The NSERIES \cite{2017MNRAS.470.2617A} mocks are a suit of high-resolution N-body simulations that were used in both BOSS DR12 and eBOSS DR16 analysis. Their main purpose was to test the various fitting methodologies used for theoretical systematics. NSERIES consists of 84 mock catalogues. 

These mocks are generated from seven independent simulations conducted in a volume of $(2.6h^{-1} \text{ Gpc})^3$ and created using the $N$-body code \texttt{GADGET2} \cite{Springel:2005mi}. Furthermore, each simulation is projected into seven different orientations and cuts, resulting in a total of 84 distinct mock datasets. These mocks are populated with galaxies using an HOD scheme designed so that the galaxy catalogue matches the CMASS sample. The  cosmological parameters adopted for N-Series are: $\Omega_{m}=0.286$, $h=0.7$, $\Omega_{b}=0.047$, $\sigma_{8}=0.820$, $A_s \times 10^9=2.146$  and $n_{s}=0.96$.  Here, we use the cutsky NSERIES mocks, whose footprint and number density correspond to that of CMASS north galactic cap at a redshift of z=0.55.

\item The MultiDark Patchy BOSS DR12 mocks (hereafter MD-Patchy mocks) \citep{Kitaura2014,Kitaura2016} are a suit of 1000 simulations used to estimate the covariance matrix required for analyzing BOSS data. MD-Patchy mocks are based on second-order Lagrange perturbation theory and use a stochastic halo biasing scheme calibrated on high-resolution N-body simulations. Each mock is built from a box of $(2.5h^{-1} \text{ Gpc})^3$ and is populated with halos following an HOD scheme calibrated to match the BOSS samples. The MD-Patchy cosmology is: $\Omega_{m}=0.307115$, $\Omega_{\Lambda}=0.692885$, $\Omega_{b}=0.048$, $\sigma_8=0.8288$ and $h=0.6777$. MD-Patchy mocks were designed to match the number density and footprint of both the CMASS and LOWZ samples from Data Release 12 and were also split into the 3 redshift bins defined above.

\end{itemize}

\section{Modelling the redshift space correlation function}
\label{sec:model}
In this work we adopt a Lagrangian approach, on which we follow the trajectories of cold dark matter particles with initial position $\vq$ through the map
\begin{equation}\label{qTox}
    \vx(\vq,t) = \vq + \Ps(\vq,t),
\end{equation}
where $\Ps$ is the Lagrangian displacement field.
The observed positions of objects are distorted by the Doppler effect induced by their peculiar velocities relative to the Hubble flow, $\vv = a \dot{\vx} = a \mathbf{\dot{\Psi}}$. That is, for a tracer located at a comoving real space position $\vx$, its apparent redshift-space position becomes $\vs= \vx + \vu$, with along the line-of-sight ``velocity'' $\vu$ 
\begin{equation}
\vu \equiv  \vhn \frac{\vec v \cdot \vhn}{a H} = \vhn \frac{\mathbf{\dot{\Psi}} \cdot \vhn}{H},
\end{equation}
where we are using the distant observer approximation on which the angular observed direction of individual galaxies $\hat{\vx}_i$ are replaced by a single line-of-sight direction $\vhn$, which is representative to the sample of observed objects. 
The map between Lagrangian coordinates and redshift-space Eulerian positions becomes 
\begin{equation}
 \vec s = \vq + \mathbf{\Psi} + \vhn \frac{\mathbf{\dot{\Psi}} \cdot \vhn}{H}.
\end{equation}

The correlation function of tracer counts in redshift space is given by the standard definition
\begin{equation}
    1+ \xi_s(\vs) = \langle \big(1+\delta_s(0)\big) \big(1+\delta_s(\vs)\big) \rangle
\end{equation}
where the tracer fluctuation in redshift space is $\delta_s(\vs)$. Now, the conservation of number of tracers $X$, that reads $\big[1+\delta_s(\vec s)\big]d^3s = \big[1+\delta_X(\vx)\big]d^3x$, yields the redshift-space correlation function \cite{Scoccimarro:2004tg}
\begin{equation}
  1 + \xi_s(\vec s) = \int \Dk{k} d^3r \,e^{i\vk\cdot(\vec s - \vr)} \Big[ 1+\mathcal{M}(\vk,\vr) \Big], 
\end{equation}
where $\vr = \vx_2 - \vx_1$ and $\vs = \vs_2 - \vs_1$. The density weighted pairwise velocity generating function is  
\begin{equation}
1+\mathcal{M}(\vec J,\vr) =  \left\langle \big(1+\delta_X(\vx_1) \big)\big(1+\delta_X(\vx_2)\big)  e^{-i \vec J \cdot \Delta \vec u}  \right\rangle,
\end{equation}
where $\Delta \vec u = \vec u(\vx_2)-\vec u(\vx_1)$. 

The generating function is now expanded in cumulants $\mathcal{C}$. That is, 
\begin{equation}
 \mathcal{Z}(\vec J,\vr)\equiv \log\Big[ 1+\mathcal{M}(\vec J,\vr) \Big] = \sum_{n=0}^\infty \frac{(-i)^n}{n!}J_{i_1}\cdots J_{i_n} \mathcal{C}^{(n)}_{i_1\cdots i_n}(\vr)
\end{equation}
where in the second equality we use the Taylor series of $\mathcal{Z}$ about $\vec J=0$. The cumulants are then obtained by
\begin{equation}
\mathcal{C}^{(n)}_{i_1\cdots i_n}(\vr) = i^n\frac{\partial^n \mathcal{Z}(\vec J, \vr)}{\partial J_{i_1}\cdots\partial J_{i_n}}\Bigg|_{\vec J=0}. 
\end{equation}
  Then, 
\begin{equation} \label{xisC}
  1 + \xi_s(\vs) = \int \Dk{k} d^3x \,e^{i\vk\cdot(\vs - \vr)} \exp \left[ \sum_{n=0}^{\infty} \frac{(-i)^n}{n!} k_{i_1}\cdots k_{i_n} \mathcal{C}^{(n)}_{i_1\cdots i_n}(\vr) \right], 
\end{equation}

On the other hand, the generating function can be alternatively expanded in moments $\Xi^{(n)}$ as follows
\begin{align}
 \Xi^{(n)}_{i_1 \cdots i_n}(\vr)&=   
 i^n \frac{\partial^n \,}{\partial J_{i_1}\cdots J_{i_n}} \Big(1+\mathcal{M}(\vec J,\vr) \Big)\Bigg|_{\vec J=0} \nonumber\\
 &= \left\langle\big(1+\delta_X(\vx_1)\big) \big(1+\delta_X(\vx_2)\big) \Delta u_{i_1}\cdots \Delta u_{i_n} \right\rangle,
\end{align}

which lead us to relations between cumulants and moments
\begin{align}
 \mathcal{C}^{(0)}(\vr) &=\log[1 + \xi(r)],\\
 \mathcal{C}^{(1)}_i(\vr) &= \frac{\Xi^{(1)}_i(\vr)}{1+\xi(r)}\equiv v^{\hat{n}}_{12,i}, \\
 \mathcal{C}^{(2)}_{ij}(\vr) &=  \frac{\Xi^{(2)}_{ij} (\vr)}{1+\xi(r)} - \mathcal{C}^{(1)}_i(\vr) \mathcal{C}^{(1)}_j(\vr) \equiv  \hat{\sigma}^{2\hat{n}}_{12,ij} - v^{\hat{n}}_{12,i} v^{\hat{n}}_{12,j} = \sigma^{2\hat{n}}_{12,ij},
\end{align}
where we introduced the pairwise velocity along the line of sight $v^{\hat{n}}_{12,i}$ and the pairwise velocity dispersion along the line of sight moment and cumulant, $\hat{\sigma}^{2\hat{n}}_{12,ij}$ and $\sigma^{2\hat{n}}_{12,ij}$, respectively. These relations will serve us below, since moments are more directly computed from the theory than cumulants. 

Using eq.~\eqref{xisC}, the correlation function in redshift space is
\begin{align} \label{xisF}
  1 + \xi_s(\vs) 
  &= \int d^3 r \big[1 + \xi(r)\big] \int \Dk{k} \exp\Big[i\vk \cdot (\vs - \vr - \vec v_{12}^{\vhn}) -\frac{1}{2}\vk^T \bm{\sigma}^{2\vhn}_{12} \,\vk + \cdots \Big].
\end{align}

If we stop at the second order cumulant $ \bm{\sigma}^{2\vhn}_{12}$, the $k$-integral can be formally performed analytically, giving
\begin{equation}\label{xisFi}
  1 + \xi_s(\ve s) = \int \frac{d^3 r}{(2\pi)^{3/2} |\mathbf{\sigma}^2_{12}|^{1/2}}
 \big[1+\xi(r)\big] \exp \left[ -\frac{1}{2} (\ve s -\vr -\vec v_{12}^{\vhn} )^\mathbf{T} [\bm{\sigma}^{2\vhn}_{12}]^{-1}(\ve s -\vr -\vec v_{12}^{\vhn} )  \right],   
\end{equation} 
which is the {\it Gaussian Streaming Model correlation function}. 

Now, depending on how one computes the ingredients $\xi(r)$, $v_{12,i}(\vr)$ and $\sigma_{12,ij}^2(\vr)$, different methods can be adopted from here. For example: \emph{1)} Reference \cite{Reid:2011ar} computed $\xi(r)$ within the Zeldovich approximation, but the pairwise velocity ($v_{12}$) and pairwise velocity dispersion ($\sigma_{12}^2$) using Eulerian linear theory. \emph{2)} In \cite{Wang:2013hwa} Convolution Lagrangian Perturbation Theory (CLPT) is used for the three ingredients, but instead of computing $\sigma_{12}^2$, the authors computed $\hat{\sigma}_{12}^2$.  This latter reference also released a widely used code by the community.\footnote{Available at \href{https://github.com/wll745881210/CLPT_GSRSD}{github.com/wll745881210/CLPT\_GSRSD}.} 

Here, we will use the method of \cite{Uhlemann:2015hqa, Vlah:2016bcl,Valogiannis2020}, where all moments are computed using CLPT. Further, in our modeling we consider a Lagrangian biasing function $F$ that relates the galaxies and matter fluctuations through  \cite{Matsubara:2008wx,Aviles:2018thp}
\begin{equation}\label{LagrangianF}
 1+\delta_X(\vq) = F(\delta, \nabla^2 \delta) = \int \frac{d^2\mathbf{\Lambda}}{(2\pi)^2} \tilde{F}(\mathbf{\Lambda}) e^{i \mathbf{D}\cdot \mathbf{\Lambda}}. 
\end{equation}
In the second equality $\tilde{F}(\mathbf{\Lambda})$ is the Fourier transform of $F(\ve D)$, with arguments $\mathbf{D}=(\delta, \nabla^2 \delta)$ and spectral parameters $\mathbf{\Lambda}=(\lambda,\eta)$, dual to $\mathbf{D}$.
 A key assumption that we follow here, is the number conservation of tracers,\footnote{Notice the number conservation assumption of tracers is not even true for halos. However, the biasing expansion  obtained in this way is automatically renormalized and coincides with other more popular methods that introduce the biasing through the symmetries of the theory; see \cite{Desjacques:2016bnm} for a review.} from which one obtains
\begin{equation}\label{PsitodeltaX}
 1+\delta_X(\vx) = \int \Dk{k} \int d^3q e^{i \vk \cdot (\vx -\vq)} \int \tilde{F}(\mathbf{\Lambda}) e^{i \mathbf{D}\cdot \mathbf{\Lambda} - i \vk \cdot \mathbf{\Psi}},
\end{equation}
and evolves initially biased tracer densities using the map between Lagrangian and Eulerian coordinates given by eq.~\eqref{qTox}.    
Renormalized bias parameters are obtained as \cite{Matsubara:2008wx,Aviles:2018thp}
\begin{equation}\label{renormBias}
 b_{nm} = \int \frac{d\mathbf{\Lambda}}{(2\pi)^2} \tilde{F}(\mathbf{\Lambda}) e^{-\frac{1}{2} \mathbf{\Lambda}^\text{T} \mathbf{\Sigma} \mathbf{\Lambda} } (i \lambda)^n (i \eta)^m,
\end{equation}
with covariance matrix components $\Sigma_{11} = \langle \delta_{cb}^2 \rangle$, $\Sigma_{12} =\Sigma_{21} = \langle \delta \nabla^2 \delta \rangle$ and $\Sigma_{22} = \langle (\nabla^2\delta)^2 \rangle$.
We notice $b_n =b_{n0}$ are local Lagrangian bias parameters, and $b_{\nabla^2\delta} = b_{01}$ is the curvature bias.  In this work we consider only $b_1$ and $b_2$. However, tidal Lagrangian bias, $b_{s^2}$, can be easily introduced following \cite{Vlah:2016bcl}.\footnote{Indeed, our code \texttt{gsm-eft} consider tidal bias, but we use the formulae presented in \cite{Valogiannis:2019nfz}, which differ slightly from that in \cite{Vlah:2016bcl}.}

To obtain the cumulants we need to compute the moments. 
The procedure is exactly the same as with the correlation function (the zero order moment), but now we have to keep track of the velocity fields. 
That is, using 
\begin{displaymath}
 1+\delta_X(\vx_1) = \int d^3q_1 \Dk{k_1} \frac{d\lambda_1}{2\pi} \tilde{F}(\lambda_1) e^{i\lambda\delta_{1} + i \vk_1 \cdot(\vx_1-q_1-\Psi_1)},
\end{displaymath}
we obtain
\begin{align*}
 \Xi^{(n)}_{i_1 \cdots i_n}(\vr) &= \hat{n}_{i_i} \cdots \hat{n}_{i_n} \int d^3q \int \Dk{k} e^{i \vk\cdot(\vq-\vr)}\int \frac{d\lambda_1}{2\pi}\frac{d\lambda_2}{2\pi} \tilde{F}(\lambda_1)\tilde{F}(\lambda_2) \\
 &\quad \qquad\qquad\qquad \times \hat{n}_{j_i}\cdots \hat{n}_{j_n}
 \Big\langle \frac{\dot{\Delta}_{j_1}}{H}\cdots \frac{\dot{\Delta}_{j_n}}{H} e^{i [ \lambda_1 \delta_1+ \lambda_2 \delta_2 + \vk\cdot\Delta ]} \Big\rangle,
\end{align*}
where we defined $\Delta_i \equiv \Psi_i(\vq_2) - \Psi_i(\vq_1)$ and used
$\Delta u_i =  H^{-1} (\hat{n}_j \dot{\Delta}_j ) \hat{n}_i$.

The real space correlation function $\xi_{X}(r)$, which corresponds to the zeroth-order moment for tracer $X$, is obtained within CLPT, \cite{Matsubara:2007wj,Carlson:2012bu,Vlah:2015sea,Uhlemann:2015hqa,Vlah:2018ygt,Aviles:2018thp},
\begin{align}\label{CLPTxi}
&1+\xi_{X}(r) = \int  \frac{d^3 q}{(2 \pi)^{3/2} |\mathbf{A}_L|^{1/2}} e^{- \frac{1}{2}(\ve r-\vq)^\mathbf{T}\mathbf{A}_L^{-1}(\ve r-\vq) } 
 \Bigg\{ 1 - \frac{1}{2} A_{ij}^{loop}G_{ij} +\frac{1}{6}\Gamma_{ijk}W_{ijk} \nonumber\\
&\quad + b_1 (-2 U_i g_i - A^{10}_{ij}G_{ij}) + b_1^2 (\xi_L - U_iU_jG_{ij}- U_i^{11}g_i) + b_2(\frac{1}{2} \xi_L^2 -U_i^{20}g_i - U_iU_jG_{ij}) \nonumber\\
&\quad - 2 b_1 b_2 \xi_L U_i g_i  + 2(1+b_1) b_{\nabla^2 \delta} \nabla^2 \xi_L 
       + b^2_{\nabla^2 \delta} \nabla^4 \xi_L \Bigg\},
\end{align}
where the matrix 
$A_{ij}(\vq) = \langle \Delta_i^{(1)} \Delta_j^{(1)} \rangle_c $, with $\Delta_i = \Psi_i(\vq_2) - \Psi_i(\vq_1)$, is the correlation of the difference of linear displacement fields for initial positions  separated by a distance $\vq=\vq_2-\vq_1$, which is further split in linear and loop pieces:
\begin{equation}\label{ALij}
 A_{ij}(\vq) = 2 \int \Dk{p} \big( 1 - e^{i\vp\cdot \vq} \big)\frac{p_i p_j}{p^4} P_L(p) =  A^L_{ij}(\vq) +  A^{loop}_{ij}(\vq). 
\end{equation}
where $P_L$ is the linear matter power spectrum. We further use the linear (standard perturbation theory) correlation function 
\begin{equation}
\xi_L(q) = \int \Dk{p} e^{i \vp \cdot \vq}   P_L(p),
\end{equation}

and the functions

\begin{align}
W_{ijk} =  \langle \Delta_{i} \Delta_i \Delta_{k}\rangle_c, \qquad
A_{ij}^{mn} = \langle \delta^{m}(\vq)\delta^{n}(0)\Delta_i \Delta_i \rangle_c, \qquad
U^{mn}_i = \langle \delta^{m}(\vq)\delta^{n}(0)\Delta_i \rangle_c.
\end{align}
The involved $r$ and $q$ dependent tensors are $g_i = (\mathbf{A}_L^{-1})_{ij}(r_j - q_j)$, $G_{ij} = (\mathbf{A}_L^{-1})_{ij} - g_ig_j$, and $\Gamma_{ijk} = (\mathbf{A}_L^{-1})_{\{ij} g_{k\}} - g_ig_jg_k$. 

The first and second moments of the generating function yield the pairwise velocity
\begin{align}\label{v12}
&v_{12,i}(\ve r) = \frac{f}{1+\xi_{X}(r)}  \int  \frac{d^3 q\, e^{- \frac{1}{2}(\ve r-\vq)^\mathbf{T}\mathbf{A}_L^{-1}(\ve r-\vq) } }{(2 \pi)^{3/2} |\mathbf{A}_L|^{1/2}} 
 \Bigg\{ -g_r \dot{A}_{ri} - \frac{1}{2} G_{rs} \dot{W}_{rsi} \nonumber\\
&\quad + b_1 \left( 2 \dot{U}_i - 2 g_r \dot{A}^{10}_{ri} - 2 G_{rs} U_r \dot{A}_{si} \right) + b_1^2 \left( \dot{U}^{11}_i  - 2 g_r U_r \dot{U}_i  - g_r \dot{A}_{ri} \xi_L  \right)  \nonumber\\
&\quad + b_2 \left( \dot{U}^{20}_i - 2 g_r U_r \dot{U}_i  \right) + 2 b_1 b_2 \xi_L \dot{U}_i  + 2 b_{\nabla^2\delta} \nabla_i \xi_L   \,
 \Bigg\},
\end{align}
 and the  pairwise velocity dispersion
\begin{align}\label{s212}
&\sigma^2_{12,ij}(\ve r)= \frac{f^2}{1+\xi_{X}(r)}  \int  \frac{d^3 q \, e^{- \frac{1}{2}(\ve r-\vq)^\mathbf{T}\mathbf{A}_L^{-1}(\ve r-\vq) } }{(2 \pi)^{3/2} |\mathbf{A}_L|^{1/2}}  
 \Bigg\{  \ddot{A}_{ij} - g_r \ddot{W}_{rij} - G_{rs}\dot{A}_{ri}\dot{A}_{sj} \nonumber\\ 
&\quad   + 2 b_1 \left( \ddot{A}^{10}_{ij} -  g_r \dot{A}_{r\{i} \dot{U}_{j\}} - g_r U_r \ddot{A}_{ij} \right) 
+ b_1^2 \left( \xi_L \ddot{A}_{ij} + 2 \dot{U}_i \dot{U}_j \right)  + 2 b_2 \dot{U}_i \dot{U}_j  \,
\Bigg\},
\end{align}
with $q$-coordinate dependent correlators

\begin{align}
&\dot{A}_{ij}^{mn}(\vq) = \frac{1}{f H}\langle \delta^m_1 \delta^n_2 \Delta_i \dot{\Delta}_j \rangle, 
                         \qquad  \ddot{A}_{ij}^{mn}(\vq) = \frac{1}{f^2 H^2}\langle \delta^m_1 \delta^n_2 \dot{\Delta}_i \dot{\Delta}_j \rangle, \nonumber\\
& \dot{W}_{ijk} = \frac{1}{f H}\langle  \Delta_i  \Delta_j \dot{\Delta}_k \rangle,  \qquad \ddot{W}_{ijk} = \frac{1}{f^2 H^2}\langle  \Delta_i  \dot{\Delta}_j \dot{\Delta}_k \rangle,    \nonumber\\
& \dot{U}^{mn}(\vq) = \frac{1}{f H}\langle \delta^m_1\delta^n_2 \dot{\Delta}_i \rangle. 
\end{align}
As in the case of the \textit{undotted} $A_{ij}$ function, we have omitted to write the superscripts $m,n$ when these are zero; e.g, $\dot{A}_{ij} \equiv \dot{A}^{00}_{ij}$.

Now, consider the terms inside the curly brackets in eq.~\eqref{s212}. Taking its large scales limit ($\vq\rightarrow \infty$), we obtain 
 \begin{align}
  \{ \cdots \}\big|_{q\rightarrow \infty} 
  &= \delta_{mn} \int_0^\infty \frac{dk}{3\pi^2} \left[P_L(k) + \text{loop terms} \right],
 \end{align}
which is a non-zero zero-lag correlator. However, since perturbation theory cannot model accurately null separations, one needs to add an EFT counterterm that has the same structure.  This new contribution shifts the pairwise velocity dispersion as \cite{Vlah:2016bcl}
\begin{equation} \label{alphasigma}
\hat{\sigma}^2_{12,mn} \rightarrow \hat{\sigma}^2_{12,mn} + \sigma^2_\text{EFT} \delta_{mn}  \frac{1+\xi^\text{ZA}(r)}{1+\xi^\text{CLPT}_X(r)}. 
\end{equation}
As the separation distance $r$ increases, the ratio $(1+\xi^\text{ZA}(r))/(1+\xi_X(r))$ approaches unity, then the EFT counterterm adds as a constant shift to the pairwise velocity dispersion at large scales. That is, we can identify it with the phenomenological parameter $\sigma^2_\text{FoG}$ widely used in early literature to model Fingers of God (FoG). Comparisons for the modeling of the second moment when using the EFT parameter $\sigma^2_\text{EFT}$ and the constant shift $\sigma^2_\text{FoG}$ can be found in \cite{Valogiannis:2019nfz}  (see for example Fig.~2 of that reference where a particular example exhibits a clear improvement of the EFT over the phenomenological constant shift).  Finally, we notice our counterterm is related to that in \cite{Vlah:2016bcl} by $\sigma^2_\text{EFT} = \alpha_\sigma f^2$. 

There are others EFT counterterms 
entering the CLPT correlation function and the pairwise velocity and velocity dispersion, but they
are either degenerated with curvature bias (as is the case of $c_1^\text{EFT}$) or subdominant with respect to the contribution of eq.~\eqref{alphasigma} (see the discussion in \cite{Vlah:2016bcl}). So, in this work we keep only $\sigma^2_\text{EFT}$. 

Since this EFT parameter modifies the second cumulant of the pairwise velocity generation function, 
its effect on the redshift space monopole correlation function is small, while the quadrupole is quite sensitive to it, particularly at intermediate scales $r<40$ $h^{-1}\textrm{Mpc}$.

Now, let us comeback to eq.~\eqref{xisF}, that we \textit{formally} integrated to obtain eq.~\eqref{xisFi}.
However, notice the matrix $\bm{\sigma}^{2\vhn }_{12}$ is not invertible since $\bm{ \sigma}^{2\vhn }_{12}=\sigma^2_{12}\,\vhn \otimes \vhn$ and hence $\text{det}(\bm{\sigma}^{2\vhn }_{12})=0$. Hence in the following we will approach this integration differently that will also serves us to rewrite the resulting equation in a more common form and also more directly related with the computational algorithms in a code.
 
We decompose the vectors $\vk$, $\vs$ and $\vr$ in components parallel and perpendicular to the line of sight $\vhn$: 
\begin{equation}
\vk = k_\parallel \vhn + \vk_\perp, \quad \vk = r_\parallel \vhn+ \vr_\perp, \qquad \vs = s_\parallel \vhn +  \vs_\perp,
\end{equation}
with $\vk_\perp \cdot \vhn = 0$, and so on. 
We will use the following definitions 
  \begin{align}
   \mu &= \hat{\vr}\cdot \vhn = \frac{r_\parallel}{r}, \\
   v_{12}(r) &=  v_{12,i} \hat{r}_i, \\
   \sigma^2_{12}(r,\mu) &= \mu^2  \sigma^2_{12,\parallel}(r) + (1-\mu^2) \sigma^2_{12,\perp}(r) \nonumber\\
   &= \mu^2 (\hat{\sigma}^2_{12,\parallel} - v_{12}v_{12})  + (1-\mu^2) \hat{\sigma}^2_{12,\perp}(r), \label{s212s}
  \end{align}
 with $\hat{\sigma}^2_{12,\parallel}(r) = \hat{r}_i\hat{r}_j \hat{\sigma}^2_{12,ij}$ and $\hat{\sigma}^2_{12,\perp}(r) = \frac{1}{2}(\delta_{ij} - \hat{r}_i\hat{r}_j)\hat{\sigma}^2_{12,ij}$, and 
\begin{equation}
  \vec v^{\vhn}_{12} = \mu v_{12}(r) \vhn,\qquad \text{and} \qquad \bm{\sigma}^{2\vhn}_{12} =  \sigma^2_{12}(r,\mu) \vhn \otimes \vhn,  
\end{equation}
Then, we can split the $\vk$ integral eq.~\eqref{xisF} in parallel and perpendicular to the line-of-sight integrations,
  \begin{align*} 
  \int \Dk{k} e^{i\vk \cdot (\vs - \vr - \vec v^{\vhn}_{12}) -\frac{1}{2}\vk^T \bm{\sigma}^{2\vhn}_{12} \vk } 
  &= \int \frac{dk_\parallel}{2\pi} 
  e^{ik_\parallel (s_\parallel - r_\parallel - \mu v_{12}) -\frac{1}{2} k_\parallel^2 \sigma^2_{12}} \int \frac{d^2k_\perp}{(2\pi)^2} e^{i \vk_\perp \cdot (\vs_\perp - \vr_\perp)} \\ 
  &=  \frac{e^{-\frac{1}{2 \sigma^2_{12}}(s_\parallel - r_\parallel - \mu v_{12})^2}}{[2\pi \sigma^2_{12}]^{1/2}} \dD(\vs_\perp - \vr_\perp), 
  \end{align*}
obtaining a Dirac delta function from the integral of the perpendicular component $\vk_\perp$ and a Gaussian kernel from the parallel $k_\parallel$ one.

Hence, the correlation function within the GSM becomes
 \begin{equation}\label{xisGSM}
1 + \xi_s(s_\parallel,s_\perp) = \int_{-\infty}^\infty \frac{dr_\parallel}{\big[2\pi \sigma^2_{12}(r,\mu)\big]^{1/2}}
  \big[1+\xi(r)\big] \exp \left[-\frac{\left(s_\parallel - r_\parallel - \mu v_{12}(r) \right)^2 }{2 \sigma^2_{12}(r,\mu)} \right], 
 \end{equation}
  with $r^2 = r_\parallel^2 + s_\perp^2$. This is a wide popular expression, but remind that here  $\sigma^2_{12}$ is the second cumulant of the density weighted velocity generating function, instead of its second moment. Also, it suffers correction from EFT counterterms. 
  
The streaming models \cite{1983ApJ...267..465D,Fisher:1994ks,Scoccimarro:2004tg}  describe how the fractional excess of pairs in redshift space $1 + \xi_s(\vs)$ is modified with respect to their real-space counterpart $1 + \xi(r)$: 
 \begin{equation}
  1+\xi_s(s_\parallel,s_\perp) = \int_{-\infty}^\infty dr_\parallel \big[ 1 + \xi(r) \big] \mathcal{P}(s_\parallel-r_\parallel|\vr).
 \end{equation}
Here $r^2 = r^2_\parallel + r^2_\perp$ and $s_\perp = r_\perp$.  The above expression is exact; see eqs.(1)-(12) of \cite{Scoccimarro:2004tg}. This means that a knowledge of the form of the pairwise velocity distribution function $\mathcal{P}(v_\parallel|\vr) = \mathcal{P}(s_\parallel-r_\parallel|\vr)$ at any separation $\vr$, yields a full mapping of real- to redshift-space correlations. In the GSM approximation, the distribution function becomes Gaussian centered at $\mu v_{12}$ and with width equal to  $\sigma_{12}$. 

The main drawback of this approach is that the $\mathcal{P}(v_\parallel|\vr)$ is, of course, not a Gaussian \cite{Scoccimarro:2004tg,Reid:2011ar,Bianchi:2016qen}. In \cite{Bianchi:2016qen}, the authors extract the ingredients of the pairwise velocity
disribution moments directly from simulations and use them to obtain the correlation function
multipoles using the GSM and the Edgeworth streaming model of \cite{Uhlemann:2015hqa}, finding good agreement with the redshift space correlation function extracted from the same simulations, but only above scales of around $s=20 \, h^{-1} \text{Mpc}$. Our findings also indicate that our modeling and pipeline fits well the simulations above this same scale.

\subsection{\texttt{gsm-eft} code}\label{subsec:code}  

Together with this work, we release the \texttt{C}-language code \texttt{gsm-eft}\footnote{Available at \href{https://github.com/alejandroaviles/gsm}{github.com/alejandroaviles/gsm}}, which computes the multipoles of the one-loop GSM two-point correlation function in about half a second. The code receives as input, the linear power spectrum, as obtained from \texttt{CAMB}, as well as the set of nuisance parameters: These includes the biases $b_1$, $b_2$, $b_{s^2}$ and $b_{\nabla^2\delta}$, the EFT parameters $\sigma_\text{EFT}^2$ (a.k.a. $\sigma_\text{FoG}^2$), $c_1^\text{EFT}$ and $c_2^\text{EFT}$, and the cosmological parameter $\Omega_m$, which is necessary to calculate the growth rate $f$ at the output redshift.  

Notice that the CLPT integrals involve the $\vq$-integration with a Gaussian kernel centered at $\vq=\vr$. This can be challenging for large $r$ because a naive calculation with the origin centered at $\vq=0$ will require a very fine grid for the angular integration, which should get finer as  $r$ gets larger. Hence we adapt the integrals to be always be centered at $\vq=\vr$. This change of variable allows us to perform the angular integration with high accuracy using a Gauss-Legendre method with only \texttt{gsm\_NGL}=16 weights. 

Finally, when exploring the parameter space in an MCMC algorithm, the cumulant $\sigma_{12}^2$ can become negative. To avoid this unphysical behavior we do the following.\footnote{In \cite{Vlah:2016bcl} it is warned that this can happen, and advise to keep only the linear part of $\sigma_{12}^2$ in the exponential and expand the rest. This approach is well physically motivated since only the loop terms are expanded, which is also in the spirit of CLPT. However, we indeed tried this method, and find no very satisfactory results. A second approach we followed, yielding even worst results, is to simply impose a sharp minimum cut to $\sigma^2_{12}$ to a very small, but still positive number.}  We split the cumulant of eq.~\eqref{s212s} 
 as $\sigma^2_{12}=\sigma^2_{12,L}+\sigma^2_{12,\text{loop}}$, that is, in linear and loop pieces, the latter containing the EFT counterterm and velocity moments. When $\sigma^2_{12,\text{loop}}< - c_{tol} \, \sigma^2_{12,L} $, with  $c_{tol}$ close but below unity, we transform the variable
 \begin{equation}
 \sigma^2_{12} = \sigma^2_{12,L}+\sigma^2_{12,\text{loop}} \quad \longrightarrow \quad \sigma^2_{12} = \sigma^2_{12,L}+f(\sigma^2_{12,L},\sigma^2_{12,\text{loop}})   
 \end{equation}
 with
 \begin{align}
 f(\sigma^2_{12,L},\sigma^2_{12,\text{loop}}) =  \frac{A  \, \sigma^2_{12,\text{loop}}}{\sigma^2_{12,\text{loop}}+B} -A -  \sigma^2_{12,L},    
 \end{align}
 where the constants $A$ and $B$ are given by 
 \begin{align}
     A &=  \frac{(-1 + c_{tol}  )(B-c_{tol} \, \sigma^2_{12,L})}{B} \sigma^2_{12,L}\\
     B &= (-1 + 2 c_{tol}  )\sigma^2_{12,L}
 \end{align}
 with this transformation the range $(-\infty,-c_{tol} \,\sigma^2_{12,L})$ is shortened to $(-\sigma^2_{12,L},-c_{tol} \, \sigma^2_{12,L})$, while the one-loop cumulant $\sigma^2_{12}$ stays smooth and is strictly positive. After preliminar tests, we chose the value $c_{tol}=0.999$.

\section{Accelerating Modeling with Neural Networks}
\label{sec:NN_definition}

Our full shape analysis requires an exploration of a relatively large parameter space. Each model requires approximately 1.5 seconds in our computer to run. Given the large number 
of evaluations required to explore the parameter space (of the order of $10^5$), 
 and the large amount of MCMC chains we are interested in running, there is an incentive to optimize the evaluation process of our model. 

There are various methodologies available for accelerating the estimation of these statistics. The choice between using one or another depends on several factors, one of them being the number of models that can be constructed to use as a training set. In our case, the Gaussian streaming model presented in Section \ref{sec:model} is relatively cost-efficient. To run our model, we first construct a template of the power spectrum using the publicly available Code for Anisotropies in the Microwave Background (CAMB) \citep{Lewis:1999bs,Lewis:2002ah}, which completes in approximately one second. We then utilize this template as input for our \texttt{gsm-eft} code, which requires an additional half-second to compute the correlation function multipoles.  Considering this, we can efficiently generate training data sets of several tens of thousands of points within a reasonable computational time. Recently, neural networks, have proved to be a suitable framework to accelerate the estimation of clustering statistics for training sets of this size \cite[e.g.][]{2022MNRAS.511.1771S,  2022JCAP...04..056D}. Moreover, neural networks are particularly efficient in data generalization, affording an almost constant reliability over the full parameter space (i.e. the model error does not strongly depends on the distance with the nearest point used on the training set). 

In what follows, we present our emulating methodology. Our approach is derived from the methodology proposed in \cite{2022JCAP...04..056D}, but we have made modifications to adapt it for configuration space. The following subsection provides a detailed explanation of our methodology and highlights the specific changes we have implemented to transition into configuration space. Once our neural networks  are trained, we reduce the evaluation time needed for a single point in parameter space to around  0.015 seconds, which improves in two orders of magnitude the Likelihood evaluation time. 

We note that a distinct neural network emulator is necessary for each multipole of the correlation function at a specific redshift.\footnote{One can decide to train a global neural network including the two multipoles. However, it corresponds to expanding the exit layer by a factor of three without a win of information between them.} Throughout this study, we utilize the first two non-zero multipoles of the correlation function. Consequently, each analysis presented in this work entails constructing two neural network emulators. The construction process for each emulator takes approximately 30 minutes when performed on our personal laptops. It is worth mentioning that it might be feasible to reduce the building time of a single neural network by adjusting certain parameters, as discussed below. Nevertheless, since the number of neural networks we use is small, we find the current building time to be manageable.

Other methodologies that operate in Fourier space \cite[e.g.][]{2009ApJ...705..156H, 2022JCAP...04..056D}, aim to predict values for all wave numbers of interest, which typically comprise hundreds of points. If a brute force approach were employed, where one asks the neural network to directly predict the power spectrum, it would need to make hundreds of predictions, which would increase the time required to build the network. Therefore, methodologies utilizing Fourier space often employ techniques such as principal component analysis to address this issue.\footnote{In principal component analysis, the input power spectra matrix is divided into eigenvectors, which are dependent on the wave number, and their corresponding eigenvalues, which only rely on the cosmology. This enables an approximation of the power spectra by considering a linear combination of the most significant eigenvalues and discarding the rest, reducing the number of predictions necessary.} 

 Here, we model the correlation function from 20 $h^{-1}\,\textrm{Mpc}$ to 130 $h^{-1}\,\textrm{Mpc}$ in 22 bins with a 5 $h^{-1}\,\textrm{Mpc}$ width between each bin in redshift space distance $r$, therefore, each emulator needs to predict only 22 numbers.
 
We have found no necessity to incorporate principal component analysis as part of our methodology as Fourier space works utilize a number of principal components similar to the number of bins we employ.

To train each neural network, we generate 60,000 models distributed across the parameter space.
Out of these, 50,000 models are utilized for training, while 5,000 models constitute the validation set. The validation set is employed during the training process to test the data and determine when to decrease the learning rate of the network, as explained below. The remaining 5,000 models form our test set and are reserved for evaluating the accuracy of the methodology on unseen data so that we perform a fair assessment of the trained models.

We use Korobov sequences \citep{korobov1959approximate} to select the points in the parameter space used for building and testing our neural networks. Korobov sequences are a robust approach for generating extensive and uniform samples in large dimensional spaces. We run three distinct sequences to create the training, test, and validation sets at each distinct redshift that we model. We also make sure that the three sequences are independent and that there are no overlapping points between them. Finally, we employ our EFT-GSM model pressented in section \ref{sec:model} to calculate the multipoles of the power spectra for all 60,000 data points. These EFT-GSM multipoles are used to train our neural networks, that aim to accurately replicate the GSM predictions for a new point in parameter space.

In order to make the training of the neural network more efficient, it is convenient to keep the values of the output layer neurons in a similar range of values. Here, we use a hyperbolic sinus transformation on each of the training set multipoles for this purpose. 

We run our neural networks by adapting the public code from \cite{2022JCAP...04..056D},\footnote{Which is available at \href{https://github.com/sfschen/EmulateLSS/tree/main}{https://github.com/sfschen/EmulateLSS/tree/main}} to reflect the changes expressed above. We use the Multi-Layer Perceptron architecture suggested by them, with four hidden layers of 128 neurons each. As we discuss below, the accuracy that we obtain in our predictions is well within the precision we need and this is achieved in a manageable time. Therefore, we decided that no further optimization of the architecture to fit our particular data was necessary. When training our networks, we reduce the learning rate of the algorithm from $ 10^{-2}$ to $ 10^{-6}$ in steps of one order of magnitude and double the training batch size at every step. As suggested by several works \citep[e.g.][]{2020ApJS..249....5A,2022JCAP...04..056D}, we use the following activation function,

\begin{equation}
    a(X)=\left[\gamma+(1+e^{-\beta \odot X} )^{-1}(1-\gamma)\right]\odot X,
\end{equation}

which we found outperforms other more common activation functions like Rectified linear units. Here, $\gamma$ and $\beta$ are new free parameters of a given hidden layer within the neural network that are fitted during the training process of the network.

The algorithm decreases the learning rate when a predetermined number of training epochs have passed without any significant improvement on the accuracy of the model, this number is commonly referred to as the patience of the algorithm.\footnote{We track the mean square error (MSE) of the validation set, the algorithm records the best value found so far. When a number of epochs equal to our patience value have elapsed whiteout a better MSE being found, the algorithm switches the learning rate. Note that a larger patience allows the model more time to exit local minima, and address slow convergence issues.} The patience we use determines the time required to train our neural networks, longer patience usually leads to more accurate models (provided the model is not overfitted). The results presented in this work correspond to waiting 1000 epochs before reducing the learning rate, which, as stated above corresponds to approximately 30 minutes of training time, also, we have monitored our validation set to ensure that there are no signs of overfitting at this point.\footnote{ An overfitted model would start to worsen the MSE of the validation set after a given training epoch.} If our goal were to reduce the training time of the algorithm, we could decrease the patience value. However, this would reduce the accuracy of our models, we note that a patience of around 100 epochs reduces the training time to around 5 minutes on our personal laptops while still maintaining sub-percent accuracy in most multipole models.

\section{Methodology}\label{sec:method}

In this section, we define the methodology employed to extract cosmological information from galaxy clustering. First, we describe the clustering measurements we utilize. Next, we provide an overview of our full shape methodology. We also discuss the likelihood, priors, covariance, and MCMC samplers used throughout our analysis.

\subsection{Clustering measurements and fiducial cosmology}
\label{multipoles_data}

Throughout this work, we focus on the anisotropic 2-point correlation function $\xi(\mu,s)$, which we project under the Legendre polynomial basis $L_\ell(\mu)$ following equation \ref{multipoles}:
\begin{equation}\label{multipoles}
\xi_\ell(s)\equiv \frac{(2 \ell +1)}{2} \int_{-1}^{1} L_\ell(\mu)\xi(\mu,s)d\mu.
\end{equation}

Here, $\ell$ is the order of the polynomial and $\mu$ is the cosine of the angle between the separation vectors and the line-of-sight direction.

We use the legacy multipoles from BOSS \cite{2017MNRAS.470.2617A} computed with the fiducial cosmology $\Omega_m=0.31$, $\Omega_\Lambda=0.69$, $\Omega_k=0$, $\Omega_b h^2=0.022$, $\Omega_\nu h^2=0.00064$, $\omega_0=-1$, $\omega_a=0$, $h=0.676$, $n_s=0.97$, and $\sigma_8=0.8$. 

In this work, we utilize the first two non-zero multipoles of the correlation function that correspond to $\ell=0,2$.

\subsection{Full Shape Methodology}
\label{sec:fullshape}
The full-shape methodology followed to constrain cosmological parameters consists of varying a theoretical model $\xi_\ell^\mathrm{Model}(s)$ (or the equivalent statistics in Fourier space) at different points in parameter space and comparing the resulting models directly with the measured clustering $\xi_\ell^\mathrm{Data}(s)$ without any compression of the information. The way in which we select the points to be explored in parameter space is described in section \ref{sec:likelihoodpriors} below. Throughout this work, we use the GSM-EFT model to build templates of the multipoles of the galaxy correlation function in redshift space, using eq.~\eqref{xisGSM}. The methodology implemented here can be used with any other perturbation theory correlation function code; e.g. \texttt{Velocileptors}\footnote{\href{https://github.com/sfschen/velocileptors}{https://github.com/sfschen/velocileptors}} \cite{Chen:2020fxs,Chen:2020zjt}. In order to compute a given correlation function template with GSM-EFT, it is necessary to provide a fixed value of the free parameters of the model. These free parameters can be divided into three distinct subsets. The first subset corresponds to the cosmological  cosmological parameters required to construct the linear power spectrum from CAMB, these parameters are $h$, $\omega_{b}$, $\omega_{cdm}$, $A_s$, $n_s$, $N_\mathrm{eff}$, $\Omega_\mathrm{ncdm}$. Our second set of parameters are the three nuisance parameters used to model the relationship between galaxies and matter and the EFT counterterms. These parameters are $b_1$, $b_2$, $b_{\nabla^2\delta}$ and $b_{s^2}$, and  $\sigma_{\mathrm{EFT}}^2$ and $c_{1,\mathrm{EFT}}$.  

As explained in section 4, we use surrogate models built with neural networks to optimize the speed at which we can generate theoretical templates. Clustering measurements employ a reference cosmology for transforming redshift to distance measurements. This reference cosmology introduces Alcock-Paczyński distortions that must be considered when comparing our data and model multipoles. To address this issue, we employ a pair of late-time re-scaling parameters, denoted as $ q_{||}$ and $q_{\perp}$, which introduce the necessary corrections to the galaxy clustering in two directions: along and perpendicular to the line of sight. This approach enables us to account for the impact of an inaccurate fiducial cosmology when calculating the clustering. The  components of the separation in the true cosmology ($s_{||}'$,$s_{\perp}'$) are expressed in terms of the components of separation in the fiducial cosmology ($s_{||}$,$s_{\perp}$) as follow:
\begin{equation}
\label{AP_formulation}
s_{||}'=s_{||}q_{||},\;\;
s_{\perp}'=s_{\perp}q_{\perp}.  
\end{equation}
The geometric distortion parameters, perpendicular and parallel to the line of sight, are defined as
\begin{equation} \label{APequations}
q_{\perp}\left(z_{}\right)=\frac{D_{A}\left(z_{}\right) }{D_{A}^{\mathrm{ref}}\left(z_{}\right) }, 
\quad
q_{\|}\left(z_{}\right)=\frac{H^{\mathrm{ref}}\left(z_{}\right) }{H\left(z_{}\right) },
\end{equation}
here  $D_{A}$ is the angular diameter distance, $H$ is the Hubble parameter, and the $\mathrm{ref}$ superscript indicates that the estimate is done in the reference or fiducial cosmology of the data multipoles. We use an alternative parametrization of the distortion parameters defined as:
\begin{equation}
q_{\alpha}=q_{||}^{1/3}q_{\perp}^{2/3} \; \;q_{\epsilon}=\left ( \frac{q_{||}}{q_{\perp}} \right)^{1/3} 
\end{equation}
We implement the distortions directly in the clustering by replacing  $s\rightarrow s'(s_{\mathrm{ref}},\mu_{\mathrm{ref}})$ and $\mu\rightarrow \mu'(\mu_{\mathrm{ref}})$, this can be computed using the re-scaling parameters  $\{q_{\alpha},q_{\epsilon}\}$ as follows:

\begin{eqnarray}
s'(s_{\mathrm{ref}},\mu_{\mathrm{ref}})=s_\mathrm{ref} \; q_{\alpha}\sqrt{(1+q_\epsilon)^4 \mu_{\mathrm{ref}}^2 +(1-\mu_{\mathrm{ref}}^2)(1+q_\epsilon)^{-2}}  , \\
\quad \mu'^2(\mu_\mathrm{ref})=\left [  1+ \left (\frac{1}{\mu_{\mathrm{ref}}^2 } -1\right)(1+ q_\epsilon)^{-6} \right ]^{-1}
\end{eqnarray}

The multipoles $\xi_\ell(s_\mathrm{ref})$ are estimated in the reference cosmology with $s'(s_\mathrm{ref},\mu_\mathrm{ref}) $ and $\mu'(\mu_\mathrm{ref}) $. In order to apply the dilation parameters into our implementation, we interpolate each multipole $\xi_\ell^\mathrm{model}$ using $s'(s_\mathrm{ref},\mu_\mathrm{ref})$, we also compute the observed Legendre polynomials $\mathcal{L}^\mathrm{obs}(\mu')$ using $\mu'(\mu_\mathrm{ref})$.  

Finally we construct $\xi^\mathrm{obs}(s'(s_\mathrm{ref},\mu_\mathrm{ref}) , \mu'(\mu_\mathrm{ref}))$, as 
 the sum of the multipoles times their respective Legendre polynomial, and the \textit{observed} multipoles  in the reference cosmology become
\begin{equation}
\xi_\ell (s')= \sum_{ \ell'} a_{\ell \ell'}\xi_{\ell'}(s),
\end{equation}
As expected, when using the distortion parameters the different multipoles get mixed, and so the matrix  $a_{\ell\ell'}$ is not diagonal. In principle, since we are working up to one-loop the sum is truncated at $\ell=8$. However, notice first that the for a fixed $\ell$, the dominant coefficient is $a_{\ell\ell}$. Secondly, the loop contributions of multipoles $\ell=6$ and 8 are highly suppressed, in comparison to the one-loop contribution of ultipoles $\ell=0, 2$ and 4,  because the correlation function is a very smooth function on $\mu$ at large scales. Because of this, it is an excellent approximation to truncate the sum at $\ell'=4$.  

That is, to simplify our data analysis, we have chosen not to incorporate the dilation parameters nor their effects into our neural network training.

This choice has two advantages: first, the neural network is trained without specifying a reference cosmology, leaving the possibility of changing it in order to compare different reference cosmologies; second, training the network to reproduce the multipole vectors is more convenient than training it to reproduce the 2D correlation function. 

As we have stated, our primary objective is to determine the posterior distributions of the cosmological parameters given our data multipoles. 

These posterior distributions are found by doing a thorough exploration of the parameter space using MCMC chains. In the following section, we present the methodology we employ for this exploration.

\subsection{Likelihood and Priors}
\label{sec:likelihoodpriors}
 
Since we are not interested in model comparison, we can express the posterior distribution of a point in parameter space as:
\begin{equation}
\mathcal{P}(\boldsymbol{\theta} \mid \boldsymbol{D})\propto \mathcal{L} (\boldsymbol{D} \mid \boldsymbol{\theta}) \times \mathcal{\pi}(\boldsymbol{\theta}). 
\end{equation}

Here, $\mathcal{L} (\boldsymbol{D} \mid \boldsymbol{\theta})$ and $\mathcal{\pi}(\boldsymbol{\theta})$ are the likelihood and the prior distributions, respectively. We assume Gaussian errors on the 2-point correlation function data, and therefore, the likelihood can be written as
\begin{equation}
\label{eq:posterior}
\mathcal{P}(\boldsymbol{D} \mid \boldsymbol{\theta}) \propto  
(\chi^{2})^{\frac{\nu-2}{2}} \exp\left(-\frac{\chi^{2}}{2}\right),
\end{equation}
where $\nu$ is the number of degrees of freedom, and $\chi^{2}$ is defined as:
\begin{equation}
\chi^{2}= (\vec{m}-\vec{d})^{T} C^{-1}(\vec{m}-\vec{d}),
\end{equation}
where $\vec{m}$ and $\vec{d}$ are the model and data vectors, respectively, and $C$ is the covariance matrix of the data. 

Our sample covariance matrix, between bins $i$ and $j$, is computed from the 1000 MD-Patchy mock realizations, as presented in section \ref{subsec:simulations}, using the following expression:

\begin{equation}
\begin{aligned}
C_{s}^{(ij)}=\frac{1}{N_\mathrm{mocks}-1}\sum_{m=1}^{N_{\mathrm{mocks}}}\left(\xi_i^m-\bar{\xi}_i\right)\left(\xi_j^m-\bar{\xi}_j\right)
\end{aligned}
\end{equation}

where $N_{\text{mocks}}$ represents the number of mock samples, and $\bar{\xi}_i$ denotes the average of the $i^{th}$ bin in the analysis.
We also include the Hartlap corrections \cite{Hartlap:2007}, which involve rescaling the inverse sample covariance matrix as 
\begin{equation}
\begin{aligned}
C^{-1}=C_s^{-1} \frac{N_{\mathrm{mocks}}-N_{\mathrm{bins}}-2}{N_{\mathrm{mocks}}-1}.
\end{aligned}
\end{equation}

Our parameter space exploration is done using \texttt{emcee} \cite{Foreman-Mackey:2012any}, an open MCMC code that implements the affine invariant ensemble sample proposed in \cite{emceetheory}. The boundaries of the regions explored are delineated by a set of predefined priors, which are presented in Table \ref{tab:gsm_priors}. 
As shown in the table, we explore seven parameters and held the remaining parameters constant. 
 Almost all of our parameters are assigned flat priors that correspond to the boundaries of the hyperspace within which our neural network is trained. We have checked that these boundaries are sufficiently large so the priors can be considered uniform. The only exception is $\omega_{b}$, for which we have employed a Gaussian prior. 
 
For this work we decided to use a local Lagrangian bias prescription, which means to fix $b_{s^2}$ and $b_{\nabla^2\delta}$ to zero. Further, since $c_{1,\mathrm{EFT}}$ is highly degenerate with higher-derivative bias, we also keep it fixed to zero. Hence, the only nuisance parameters considered in this work, are $b_1$, $b_2$ and $\sigma^2_{EFT}$. Further, as we show in the upcoming sections, using this simplification we can recover the cosmological parameters of simulated data with high accuracy, and our posteriors when fitting the BOSS DR12 correlation function are competitive with other analyses of the full-shape power spectrum in the literature.\footnote{In a work currently in preparation (Sadi Ramirez et al, \textit{In prepararion}), which compares compressed and full-shape methodologies, we will relax these assumptions  } We observe that our approach is not a complete one-loop theory because of the constraints on the free parameters. Consequently, we expect that the posterior distributions we have obtained would be more extensive if all parameters were unrestricted. However, it is worth noting that the existing full-shape studies in the literature rely on Gaussian priors for certain biasing, EFT, or shot noise parameters. Therefore, they also do not constitute full one-loop analyses. 

In table \ref{tab:gsm_priors} we show the varied parameters and their priors for our baseline full-shape analyses. We keep fixed the slope $n_s=0.97$, the effective number of relativistic degrees of freedom $N_\mathrm{eff}=3.046$, and the massive neutrino abundance $\omega_\mathrm{ncdm} = 0.00064$. 

\begin{table}
\ra{1.4}
\centering                          
\begin{tabular}{lc}        
\hline
\multicolumn{2}{c} {Free parameters and priors} 
\\
\hline

\underline{\texttt{Cosmological}}& \\
$\, h$ &
$\mathcal{U}[0.55, 0.91]$ \\
$\, \omega_{b}$&
$\quad \mathcal{N}[0.02237,0.00037]\quad$ \\
$\, \omega_{cdm}$ &
$\mathcal{U}[0.08,0.16]$ \\
$\, \log(10^{10} A_s)$ &
$\mathcal{U}[2.0, 4.0]$ \\[0.2cm]

\underline{\texttt{Nuisances}}& \\
$\, b_1$ &$\mathcal{U}[0,2.0]$\\
$\, b_2$ &$\mathcal{U}[-5,10]$ \\
$\, \sigma^2_\text{EFT}$& $\mathcal{U}[-20,100]$\\[0.2cm]
\hline
\end{tabular}
\caption{Free parameters and priors we use for the baseline full-shape analyses, of both our Nseries mocks and our BOSS data. We fix the spectral index of the power spectrum to the Planck 2019 value $n_s=0.97$. We also assume a local Lagrangian biasing scheme. 
}    
\label{tab:gsm_priors} 
\end{table}

Finally, to ensure convergence, we utilized the integrated autocorrelation time, checking it at intervals of 100 steps. Convergence criteria were considered reached if two conditions were met simultaneously: the chain's length exceeded 100 times the estimated autocorrelation time, and the change in this estimation remained below the 1 per cent.

\section{Validating our Methodology with High Precision Mocks}
\label{sec:testmocks}

We have introduced our methodology for generating full-shape EFT-GSM models of the correlation function multipoles. Our ultimate objective is to re-analyze the BOSS data sets presented in section \ref{sec:data}. We will present all the results on real data in section \ref{sec:results} below. In this section, we establish a series of tests to evaluate the performance of our methodology. These tests involve applying our methodology to the NSERIES simulations presented in section \ref{subsec:simulations}.

We begin in section \ref{validation_GSM} by assessing the accuracy and precision with which our methodology can recover the parameters of the simulations. The results presented in this section are built utilising the surrogate models built with the neural network presented in section \ref{sec:NN_definition}, here we assume that these surrogates are a fair representation of the EFT-GSM models. Then, in section \ref{Testing_Neural_Networks}, we test this assumption by comparing the results of our neural network surrogate models with those from the EFT-GSM model.

\subsection{Testing EFT-GSM Model}
\label{validation_GSM}

As stated above, we assess the effectiveness of our methodology by recovering the known free parameters of the N-series simulations. In this section, we show the accuracy and precision of these results. We fit the mean multipole of the 84 mocks instead of fitting one individual mock. This approach effectively mitigates shot noise errors in the multipole models caused by inaccuracies in the shape of a single multipole.

Our error estimates are computed using one thousand MD-Patchy $z_{3}$ simulations, which are introduced in section \ref{subsec:simulations}. For estimating the sample covariance we used the multipoles from the combined sample that includes NGC and SGC which corresponds to a $V_\mathrm{eff}=4.1h^{-3}\mathrm{Mpc}^3$. We rescale the covariance matrix by a factor of $1/10$ ($10 \times 4.1 h^{-3}\mathrm{Mpc}^3$), this is done to test the methodology in a volume of the order of DESI volume, so that we can assess whether our methodology accuracy will suffice for the upcoming next-generation surveys.

Given the complexity of modelling clustering statistics at weakly non-linear scales, we should assess the scales in redshift space distance at which our model still generates accurate fits. With this in mind, we simultaneously fit both the monopole and quadrupole of the correlation function using three different ranges with varying lower limits: $s_{\text{min}} = 20, 30$ and $40 \,h^{-1}\textrm{Mpc}$. We use these fits to determine the range at which our model estimate of the parameters gets closer to the true values. Throughout this work, we maintain a fixed upper limit for our fits at $s_{\text{max}} = 130\,h^{-1}\text{Mpc}$, and we fix the width of our distance bins to 5 $h^{-1}\textrm{Mpc}$.

The resulting parameter estimations, along with the error estimates obtained through Markov Chain Monte Carlo (MCMC), are presented in Table \ref{tab:Nseriestable}. The table illustrates that, in general, the errors become narrower as the minimum scale decreases. Specifically, when employing a minimum scale of $20 \,h^{-1}\textrm{Mpc}$, we recover the most stringent constraints on all parameters and still consistent with the simulation Cosmology.

\begin{center}
\begin{table*}
\ra{1.7}
\begin{center}
\begin{tabular} { lcccccc}

Parameter               & $\,\,\quad s_{\mathrm{min}}=20$ $h^{-1}\,\textrm{Mpc} \quad $                   & $\quad s_{\mathrm{min}}=30$ $h^{-1}\,\textrm{Mpc} \quad $                    & $\quad s_{\mathrm{min}}=40$ $h^{-1}\,\textrm{Mpc} $  \\
\hline
$\omega_{cdm}$& $0.1172^{+0.0028}_{-0.0023}$ &$0.1176\pm 0.0036$          &  $0.1204\pm 0.0045$ \\
$\omega_{b}$  & $0.02305\pm 0.00036$         &  $0.02305\pm 0.00036 $      &  $0.02303\pm 0.00036$ \\
$h$          & $0.6998\pm 0.0064$            & $0.7014\pm 0.0068 $          &   $0.7036\pm 0.0072 $ \\
$\mathrm{ln}(10^{10}A_s)$ &  $3.003\pm 0.059$             & $2.986\pm 0.069$       & $2.921\pm 0.089 $    \\
$\sigma_{\mathrm{EFT}}^2$ & $14^{+13}_{-7}$          & $24^{+20}_{-20} $     &   $20^{+20}_{-20} $  \\
$b_1$          &  $1.060^{+0.083}_{-0.070}$               & $1.093\pm 0.087 $       &$1.18\pm 0.12 $  \\
$b_2$         &  $0.17^{+0.97}_{-0.65} $       & $0.9^{+1.3}_{-1.0} $         &$1.6\pm 1.8 $  \\
\hline

\bottomrule

\end{tabular}
\caption{Mean and 0.68 c.i. for all free parameters in our methodology determined using our MCMC approach when applied to the mean of the 84 NSERIES mocks. The fits are conducted simultaneously on the monopole and the quadrupole of the correlation function. The various columns display results for different values of the minimum range scale, as denoted by the column titles.
}
\label{tab:Nseriestable}
\end{center}
\end{table*}
\end{center}

\begin{figure*}
\begin{center}
\includegraphics[width=110mm]{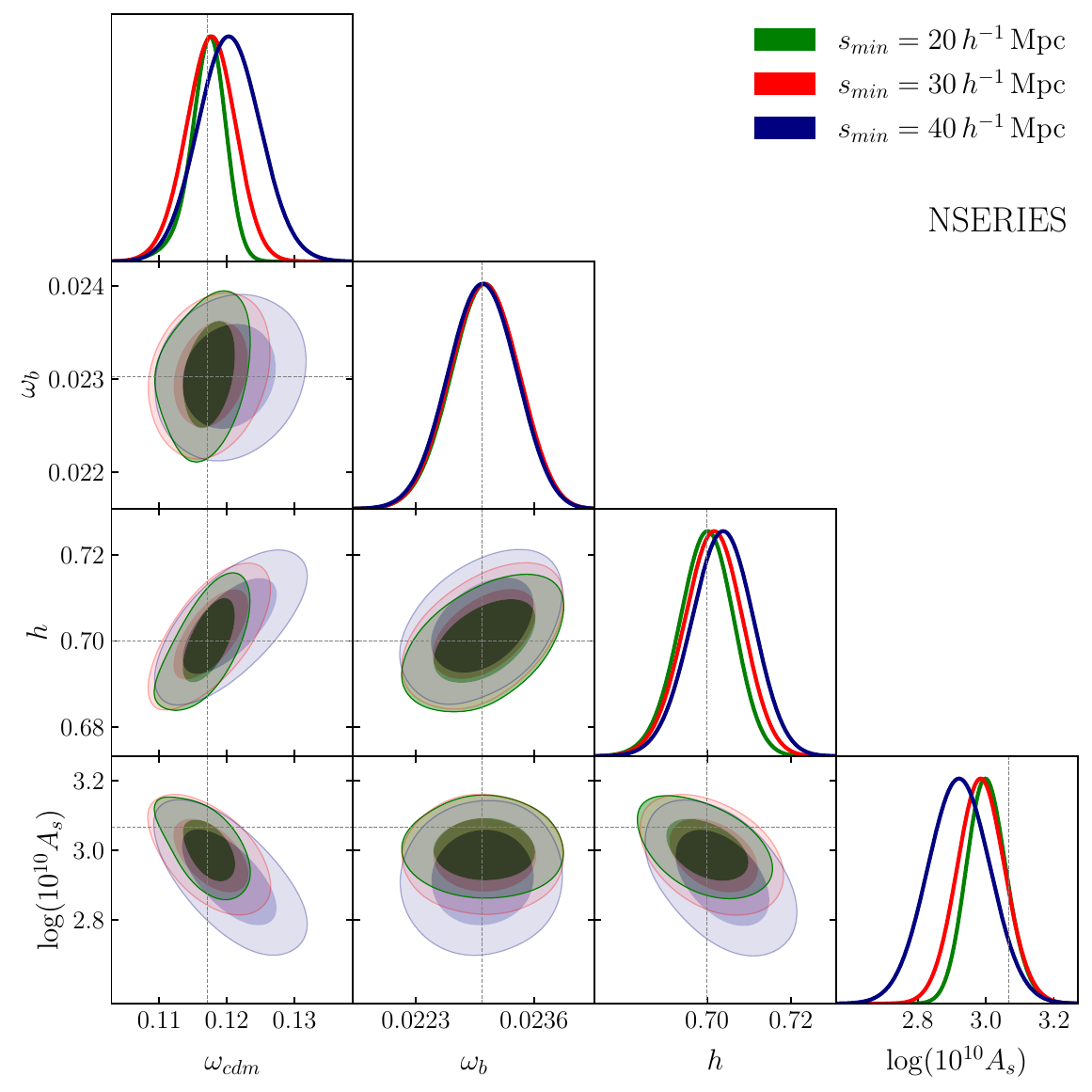}
\caption{ \textit{The effect of $s_{min}$}. Triangle plots showing the $1\sigma$ and $2\sigma$ confidence 1-dim and 2-dim regions of our cosmological parameters as obtained through our MCMC methodology. We fit to the mean of the 84 NSERIES mocks. We fit the correlation function multipoles $\ell=0$ and 2 simultaneously. The different colors denote various ranges, as specified in the figure caption. The solid gray lines depict the true parameters of the NSERIES simulations. The ranges with $s_{min} = 20\,h^{-1}\text{Mpc}$ and $s_{min} = 30\,h^{-1}\text{Mpc}$, depicted with green and red colors, respectively, exhibit similar levels of accuracy, with slightly greater precision observed in former.}
\label{Triangular_plot_Nseries}
\end{center}
\end{figure*}

\begin{figure*}
\begin{center}
\includegraphics[width=130mm]{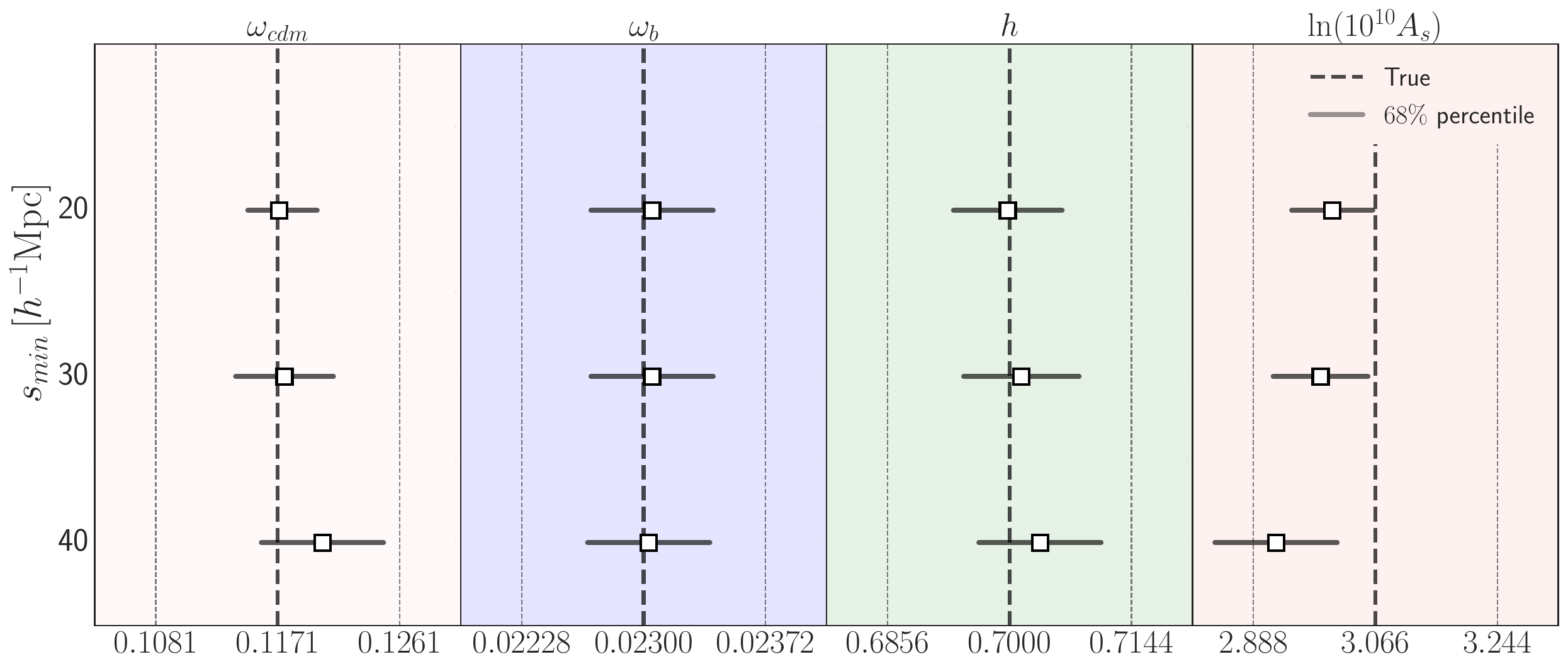}
\caption{ We present the mean value estimates obtained from our MCMC chains as squared points, which are compared to the NSERIES cosmology represented by dashed lines for all parameters of interest. The error bar associated with each point corresponds to the $1$ $\sigma$ error estimations from the chains. The square points represent, from top to bottom, the cases explored at minimum scales of 20 $h^{-1}\textrm{Mpc}$, 30 $h^{-1}\textrm{Mpc}$, and 40 $h^{-1}\textrm{Mpc}$ respectively.}
\label{error_bars_nseries}
\end{center}
\end{figure*}

We are also interested in assessing the accuracy of our models, which involves comparing the mean values obtained from our MCMC chains with the actual cosmological values from the NSERIES simulations for our four non-fixed cosmological parameters. Figure \ref{Triangular_plot_Nseries} illustrates a triangular plot of our MCMC results. In this plot, the gray lines represent the NSERIES cosmological values, while the colored histograms depict the 1D distribution of each parameter.

We observe that for all four parameters, the predictions with a minimum range of $40$ $h^{-1},\textrm{Mpc}$ perform worse than in the other two cases. This discrepancy is particularly noticeable when comparing the histograms, which appear more centered around the gray lines in the other two scenarios. This trend can be attributed to the smaller scale bins having smaller error bars, and therefore when we exclude them the overall constraining power of the model decreases.

The colored contours in the figure represent the $1\sigma$ and $2\sigma$ confidence surfaces. It is worth noting that the actual NSERIES cosmology falls within $1\sigma$ of the mean value for the $20 \, h^{-1}\textrm{Mpc}$ case, as indicated by the intersection of all gray lines within the solid green contours. Figure \ref{error_bars_nseries} summarizes this information in a more easily interpretable format. The plot demonstrates that, in general, all three models can reproduce $\omega_{cdm}$, $\omega_b$, and $h$ within $1$ $\sigma$. However, the model with a minimum scale of $40 \, h^{-1}\textrm{Mpc}$ deviates further from the true values for both $\omega_{cdm}$ and $h$. Additionally, only the the $20$ $h^{-1},\textrm{Mpc}$ model agrees with the true value of $A_s$ within $1\sigma$. It is also worth noting that the constraints on both $\omega_{cdm}$ and $A_s$ are tighter in the model with a minimum scale of $20 \,h^{-1}\textrm{Mpc}$ compared to the one with a minimum scale of $30$ $h^{-1}\textrm{Mpc}$.

Given that the $20 \, h^{-1}\textrm{Mpc}$ constraints show to be more accurate and precise than the other two cases, in the following we fix $s_{min}$ to this value.

\subsection{Testing Neural Networks}
\label{Testing_Neural_Networks}
\begin{figure*}
\begin{center}
\includegraphics[width=120mm]{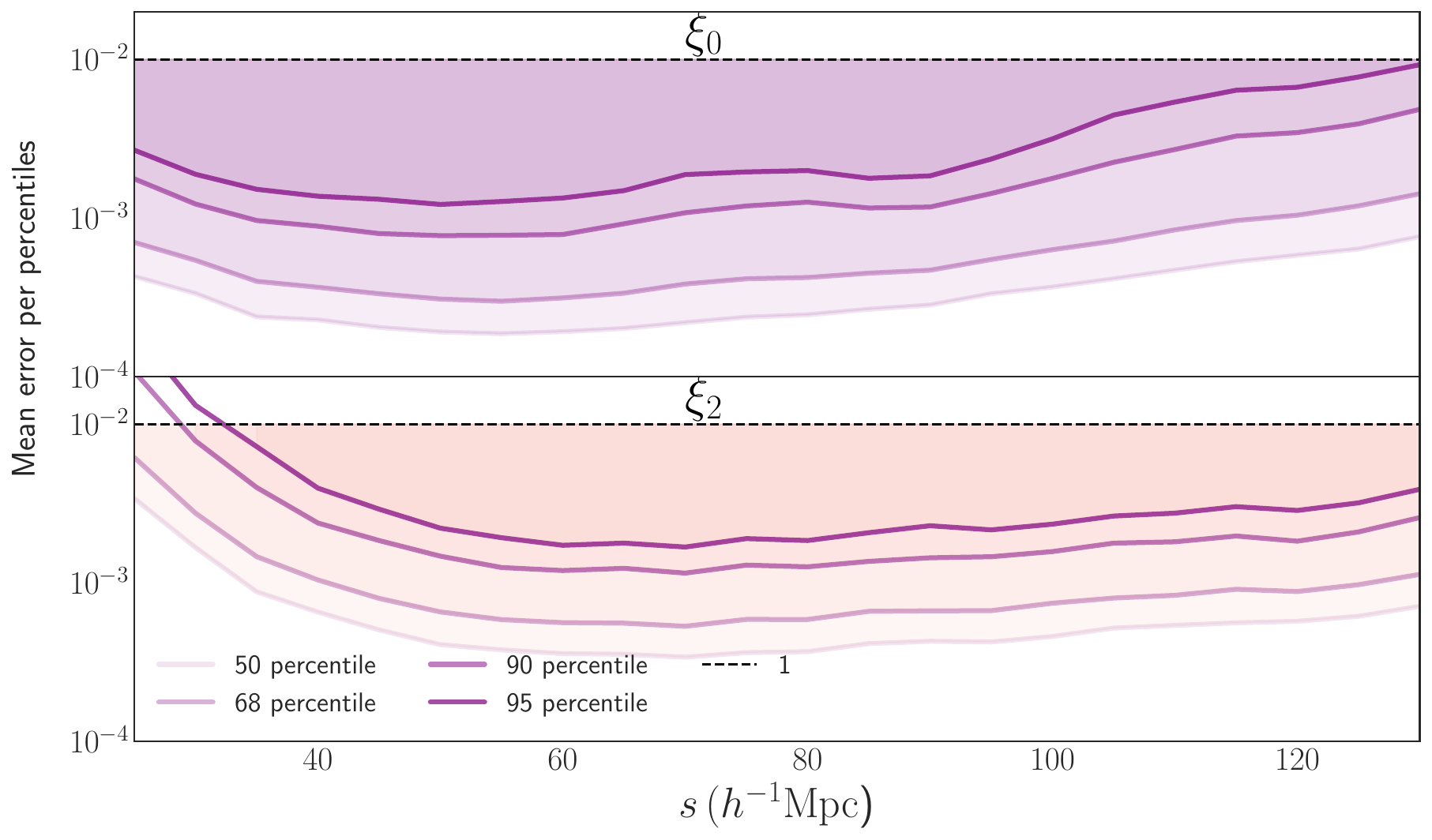}
\caption{  Percentile error at redshift $0.55$ of the predictions of our neural network for our test set, when compared to the original value of the GSM model for both the  Monopole (top) and the Quadrupole (bottom. The black line corresponds to the $1\%$ error, and the colored lines correspond to the different percentiles of errors. The plots show that at least $95\%$ of our multipoles have below percent accuracy in almost all scales and around $68\%$ of the predicted multipoles have an accuracy of around a tenth of a percent. 
}
\label{Percentile_error}
\end{center}
\end{figure*}

 In what follows we present a set of tests of the accuracy of the neural network methodology presented in section \ref{sec:NN_definition}.  We begin by testing the ability of our models to predict the multipoles of the GSM templates of section \ref{sec:model}. This is done by asking our trained networks to predict the multipoles of our 5000 test set points at redshift of $z=0.55$. For the $j^{th}$ test point, we define the percent error of the network prediction as $P_j^{err}(s)= \left| [\xi^{GSM}_{j}(s)-\xi^{NN}_{j}(s)]/\xi^{M}_{j}(s) \right|$. Where  $\xi^{GSM}_j(s)$ is the value of the multipole predicted by the GSM and $\xi^{NN}_j(s)$ is the value predicted by the neural network. This error quantifies the size of the emulator errors when compared with the size of our original statistics.

Figure \ref{Percentile_error} illustrates the percentile plots for the $50\%$, $68\%$, $90\%$, and $95\%$ percentiles of the errors. These plots show the threshold below which the specified percentage of our 5000 errors lie for a given $s$.  We note that all lines are situated below the percentile accuracy line (black line), except for the $90\%$ percentile of the quadrupole at small scales, which is only slightly above. This indicates that our neural network models are capable of reproducing the multipoles of the GSM model with an accuracy below $1\%$. Additionally, it is worth noting that the $68\%$ percentile line is positioned around the $0.1\%$ error threshold, implying that the majority of our multipoles are predicted with a precision of one-tenth of a percent, while models with an accuracy around one percent are rare.

As a second test of our methodology, we conducted two sets of different MCMC fits to the mean mock of the NSERIES from section \ref{subsec:simulations}. The first set utilizes the GSM model outlined in section \ref{sec:model}, while the second set employed a neural network surrogate model trained to replicate the behavior of the GSM model at the redshift of the NSERIES mock. We run both sets utilising two different configurations, the first is our standard range configuration of 20 $h^{-1}\,\textrm{Mpc}$ to 130 $h^{-1}\,\textrm{Mpc}$, and the second changes the minimum range to  30 $h^{-1}\,\textrm{Mpc}$.

\begin{figure*}
\begin{center} 
\includegraphics[width=75mm,height=75mm]{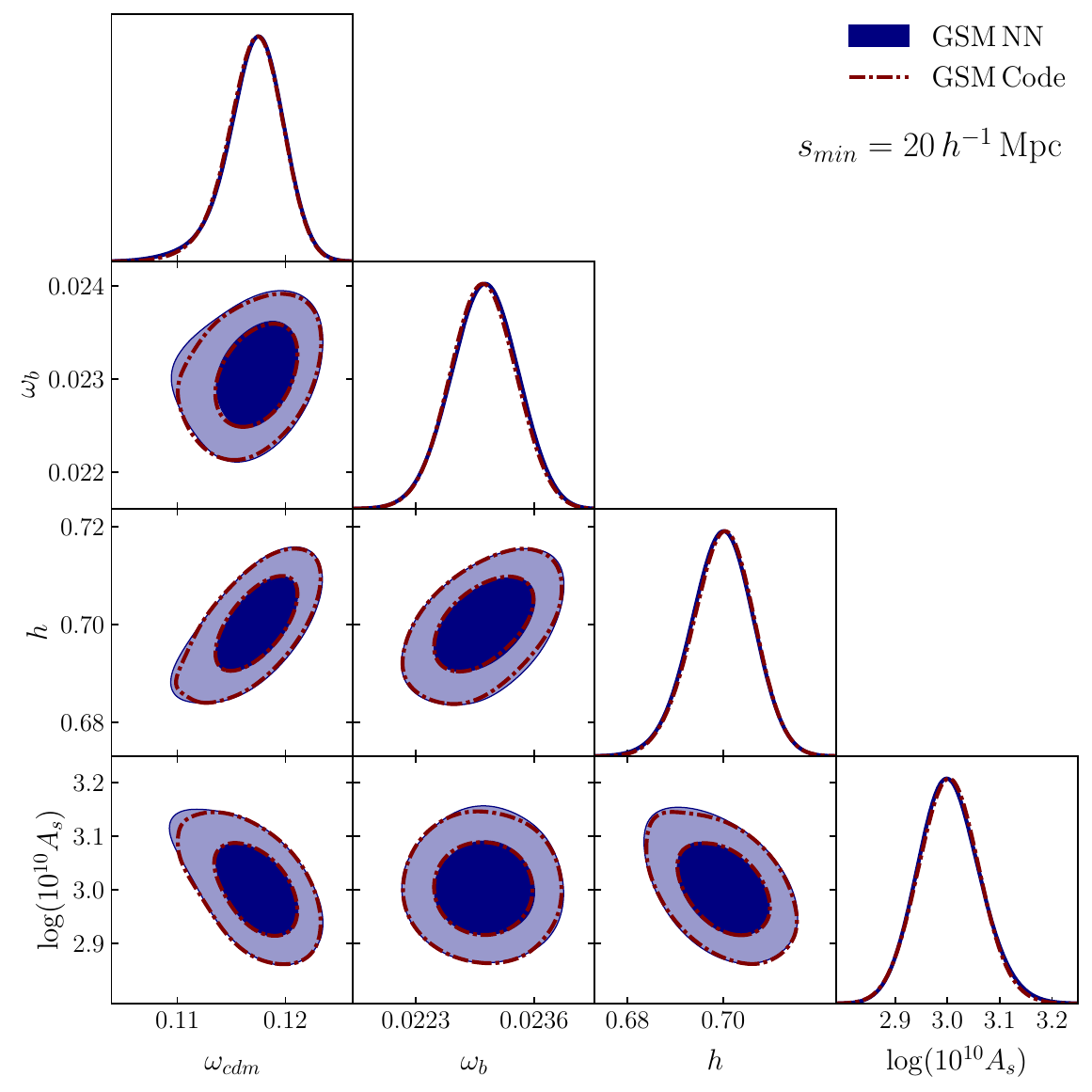}
\includegraphics[width=75mm,height=75mm]{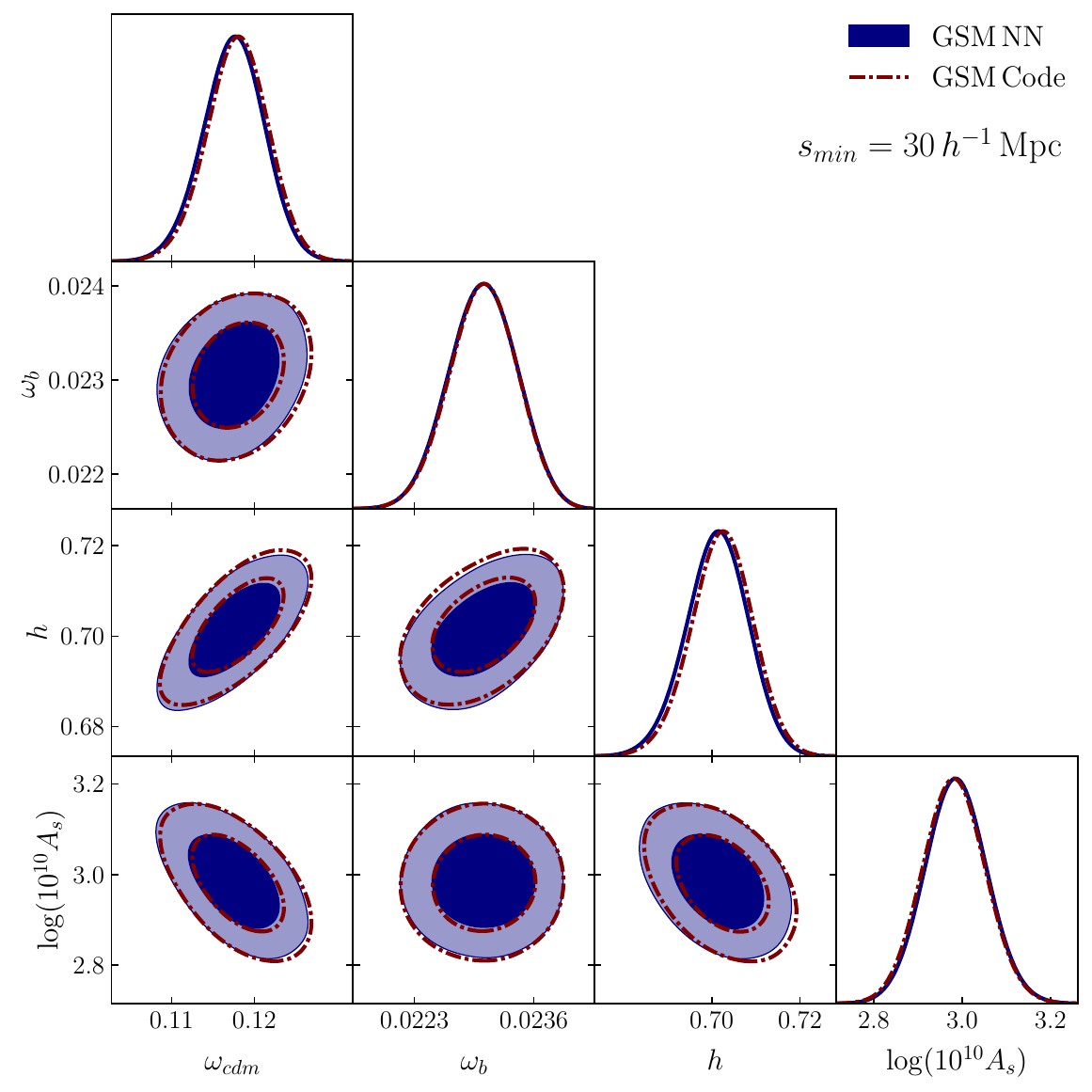}
\caption{
Triangular Plot illustrating the likelihood contours obtained by the MCMC method when we fixed to NSERIES. The solid contours represent the 1-sigma level, while the shaded contours represent the 2-sigma level. The red contours correspond to the likelihoods when training with the GSM code directly, and the blue ones when trained on our neural network model. The plot shows that both approaches yield nearly identical contours. }
\label{NN_vs_GSM_triangular}
\end{center}
\end{figure*}

Figure \ref{NN_vs_GSM_triangular} shows the triangular plots comparing the likelihood contours of both models. The 1-D histograms exhibit remarkable similarity in both plots. This results in parameter predictions that are virtually indistinguishable from each other when using the EFT-GSM model or the surrogate model. It's worth mentioning that when ussing our standard 20 $h^{-1}\,\textrm{Mpc}$ configuration there are negligible differences 
in the 2D contours that do not affect the best fits values and errors. 

Given that our neural network surrogate models can accurately reproduce the data with a significantly lower convergence time for MCMC chains, all fits presented throughout the rest of this work are built using surrogate models.

\section{Results with SDSS-III BOSS Catalogues}
\label{sec:results}

In the previous section we have shown the capability of the EFT-GSM model for recovering the cosmological parameters of the NSERIES simulations, we also tested that our surrogate models accurately reproduce the results of our EFT-GSM code. In what follows, we apply our methodology to our real galaxy data and compute our constraints on the cosmological parameters from the BOSS DR12 LRG correlation function.

\subsection{Baseline Analysis }
We begin this section by introducing our constraints on the cosmological parameters obtained by applying our baseline methodology. We computed three distinct fits, each using a different combination of the BOSS samples introduced in Section \ref{sec:data}. The first two fits utilize the monopole and quadrupole moments of the datasets $z_1$ and $z_3$ respectively. The third fit is a combined analysis where both the $z_1$ and $z_3$ multipoles were fitted simultaneously. We labeled the resulting model as $z_1+z_3$. As mentioned in section \ref{validation_GSM}, we select a scale range from $20 \, h^{-1}\mathrm{Mpc}$  to $130 \, h^{-1}\mathrm{Mpc}$ as our standard configuration. As with our NSERIES tests, our covariance matrix is computed using the MD-Patchy mocks introduced in section \ref{subsec:simulations}.

\begin{table}[h!]
\centering
\begin{tabular} { c c c c}
\hline
Parameter   & $z_1$ & $z_3$ & $z_1+z_3$\\
\hline
& & & \\[-1ex]
$\omega_{cdm}$  & $0.1038\pm 0.0064$ & $0.1238\pm 0.0076$        &$0.1115\pm 0.0050$\\[1.4ex]
$\omega_{b}$  & $0.02237 \pm  0.00037$ & $0.02236 \pm  0.00037         $  & 
 $0.02237 \pm  0.00037$ \\[1.4ex]
$h $ & $0.673\pm 0.017   $       & $0.705\pm 0.017 $  &                 $0.688\pm 0.012$\\[1.4ex]
$\ln(10^{10}A_s)$  & $3.29\pm 0.17 $  &$2.69^{+0.18}_{-0.20}           $ & $3.03\pm 0.13       $\\
\hline
& & & \\[-1ex]
$\Omega_{m}$  & $0.280\pm 0.012$ & $0.296\pm 0.015$        &$0.2846\pm 0.0093 $\\[1.4ex]
$10^{9}A_s$  &  $2.71^{+0.42}_{-0.49}$ & $1.49^{+0.23}_{-0.31}$        &$2.09^{+0.25}_{-0.29}$\\

\hline
\end{tabular}
\caption{Estimated mean and $1\sigma$ errors obtained with our MCMC methodology for three different BOSS samples $z_1$, $z_3$ and $z_1+z_3$. The fits are conducted with both the monopole and quadrupole of the correlation function. And the error covariance estimations are computed using the MD-Patchy
mocks.}
\label{tablebin_1_bin_3}
\end{table}

Table \ref{tablebin_1_bin_3} shows the constraints on our four varied cosmological parameters. We note that the error bars for $z_1$ and $z_3$ are similar, whereas the constraints for $z_1+z_3$ are slightly tighter, with $h$ and $A_s$ having error estimates that are $\sim 25\%$ smaller then the $z_1$ and $z_3$ predictions. And the errors on $\omega_{cdm}$ being around $\sim 33\%$ smaller than the one from $z_3$.

\begin{figure*}
\begin{center}
\includegraphics[width=110mm]{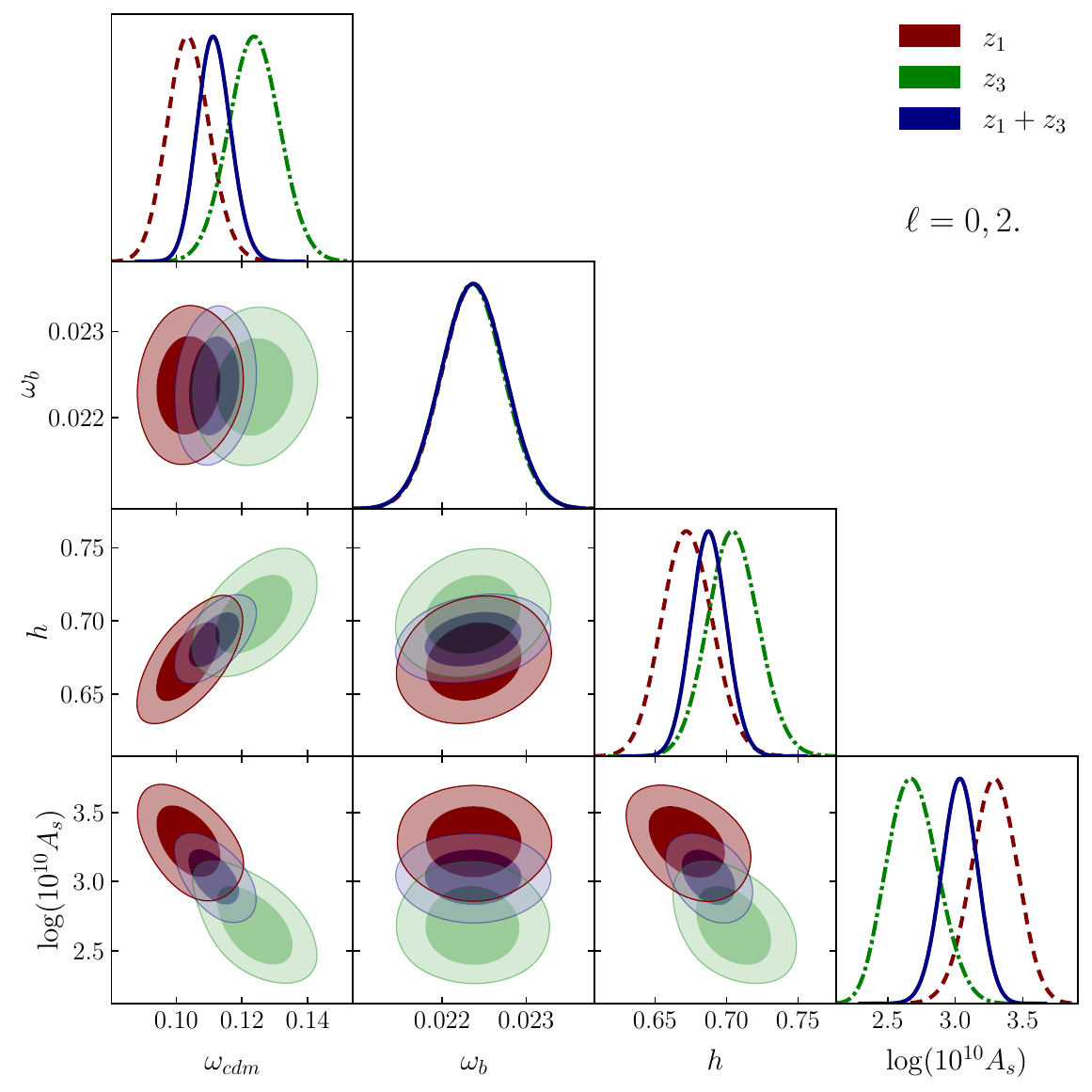}
\caption{Triangular plot showing the MCMC fits to the  BOSS data sets $z_1$ and $z_3$ and a third chain where $z_1$ and $z_3$ are fitted simultaneously, with colors as stated in the figure labels. The shaded regions correspond to the $1\sigma$ and $2\sigma$ contours and the histograms illustrate the 1D distributions of each
parameter .}
\label{Triangular plot of using bin1 and bin3 without hexadecapole}
\end{center}
\end{figure*}

Figure \ref{Triangular plot of using bin1 and bin3 without hexadecapole} shows the triangular plot of the MCMC fits to our three BOSS datasets. We observe that all parameters agree with each other within $1\sigma$, except for $A_s$, which only agrees at the $2\sigma$ level for $z_1$ and $z_3$. Mismatches between the estimated parameters for the redshift bins $z_1$ and $z_3$ are well-known, and has been reported in other works \cite{Ivanov:2019pdj,Chen_2022}, particularly in the full-shape correlation function analysis of \cite{Chen_2022}. We notice that $z_1$ predicts lower values for $\omega_{cdm}$ and $h$, while $z_3$ predicts a lower value for $A_s$. As expected, the predictions for each parameter in $z_1+z_3$ fall between the predictions from the individual samples. Notably, the predictions for $\omega_b$ are indistinguishable across all three samples as the constraints on $\omega_b$ are dominated by the prior. This is explored further in section \ref{complementary_analysis} where we widen the prior to explore the capability of LSS alone to constraint cosmological parameters and to test the methodology in a more extended parameter space.

\subsection{Comparison to other Full Shape Analysis }

As stated at the beginning of this work, several groups have reanalyzed BOSS data using a full-shape methodology. We also mentioned that most of these analyses have been conducted in Fourier space. In contrast, our work is carried out in configuration space, therefore we are interested in assessing the agreement between these two different methodologies. In this section, we compare the parameter estimations obtained from our configuration space model with a set of Fourier space results (D'Amico \citep{DAmico:2019fhj}, Ivanov \citep{Ivanov:2019pdj}, Philcox \citep{Philcox_2020}, Troster \citep{Troster_2020}, and Chen \citep{Chen_2022}), we also compare with the 
configuration space results from Zhang \citep{Zhang2022}.
To ensure a fair comparison we exclusively consider Zhang constraints derived using BOSS data, without incorporating information from other observations. As stated above an alternative to full-shape analysis is to expand the parameter space of the compression methodology by introducing a  small subset of new free parameters that account for the slope of the power spectrum. The Shapefit methodology, as presented in Brieden \citep{brieden_2021b}, employs this approach to reanalyze the BOSS data, their methodology is also developed in Fourier space. In this section, we also compare the parameter estimations obtained using our model to those obtained using the Shapefit method.

All of the analyses mentioned so far were carried out on the BOSS DR12 data, with most of them analyzing the data by dividing it into the $z_1$ and $z_3$ samples we have utilized. The only exception is D'Amico \citep{DAmico:2019fhj}, who employ the LOWZ and CMASS samples instead. Since all these studies investigate the same dataset, and the majority of them use the same samples from this dataset, we expect the parameter estimations to be consistent with each other within the uncertainty inherent to each methodology.
\begin{figure*}
\begin{center}
\includegraphics[width=125mm]{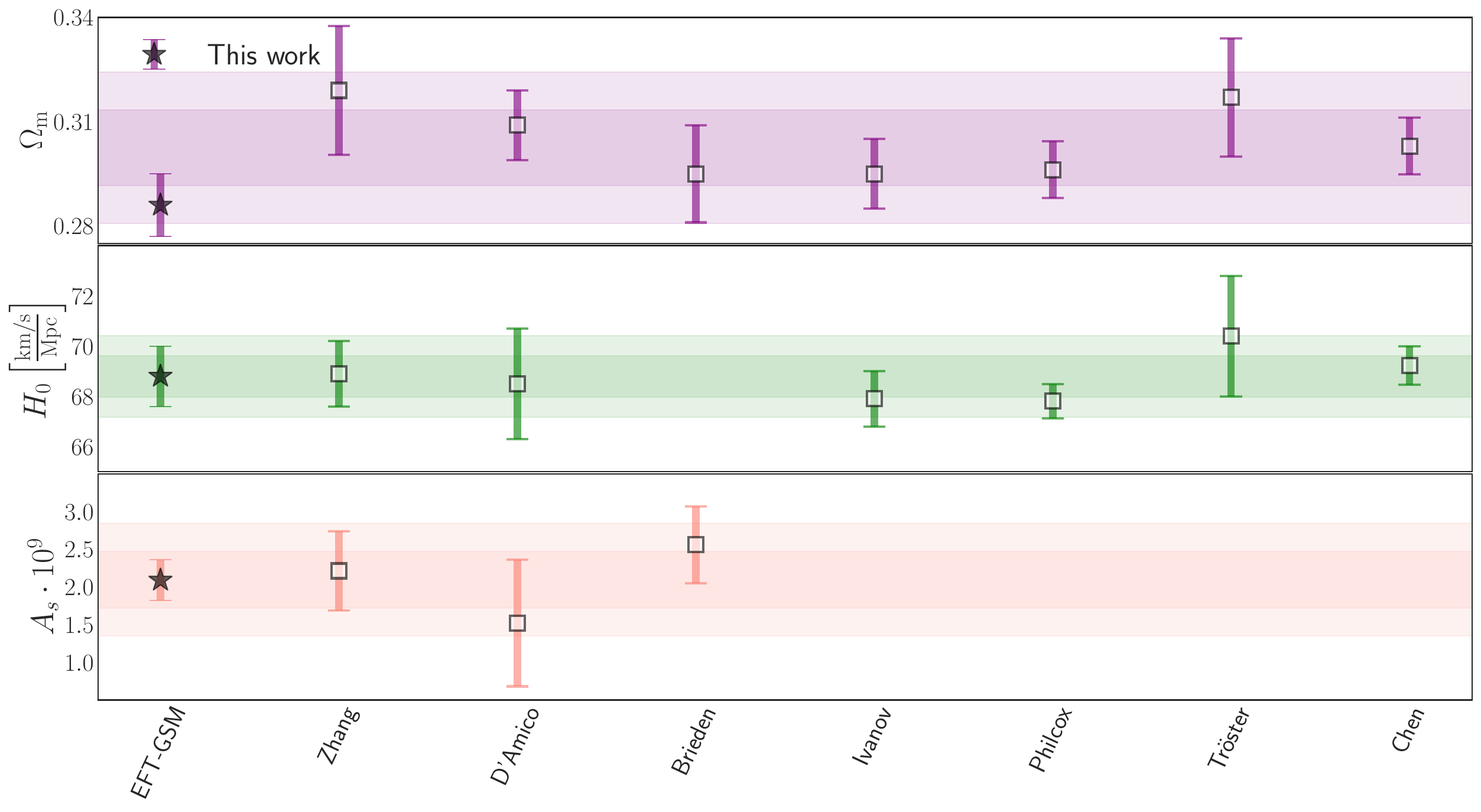}

\caption{
The starred dots represent parameter estimations obtained from BOSS $z_1+z_3$ data using our methodology. We compare these estimations with those from five other studies that perform a full shape analysis in Fourier and configuration space, as well as with the predictions from the Shapefit working group. The results from all of these other studies are denoted as square dots. The error bars associated with each data point are derived from the 1-sigma estimates obtained through MCMC analysis in each respective study. The stripes represent the 1 and 2 standard deviations around their mean, of the data shown on the different panels. }
\label{comparison_b13_papers}
\end{center}
\end{figure*}

The parameter estimations from these methodologies are depicted as square markers in Figure \ref{comparison_b13_papers}, the first column of the figure presents our parameter estimations for comparison, indicated by starred markers. We present results for three key parameters: $H_0$, $A_s$ and the total mass density $\Omega_{\mathrm{m}}$, which includes mass energy density from all matter sources, including dark matter and baryons. These specific parameters were selected to facilitate the comparison with the other works. We highlight that only two works we are comparing with include $A_s$ in their reports. Our results using these derived parameters are displayed in Table \ref{tablebin_1_bin_3}.

Figure \ref{comparison_b13_papers} shows that our predictions for both $A_s$ and $H_0$ are consistent within $1\sigma$ with the results of other studies. We also note that our predictions of $\Omega_m$ agree within $1\sigma$ with all results, except for three: D'Amico \citep{DAmico:2019fhj},  Zhang \citep{Zhang2022} and Tröster \citep{Troster_2020}, with whom we agree within  $2\sigma$. We point out that D'Amico utilises the LOWZ and CMASS samples, instead of the $z_1$ and $z_3$ samples that we use, these samples are at slightly different redshifts and use different subsets of the BOSS galaxy sample. Which should contribute to the disagreement between our measurements. Tröster employs a wide prior on the parameter $n_s$, which remains constant throughout our standard methodology. Varying this parameter has an impact on the fitting results for $\Omega_m$ and $H_0$, which should contribute to our slight disagreement.  For Zhang,  the difference observed could be explained by the two extra parameters they varied,  $n_s$ and $\Sigma m_\nu$, which can explain the difference in error bars and the position of the mean.
In Section \ref{complementary_analysis} below, we explore the effects of varying $n_s$ on our methodology. We show that when this parameter is left unfixed, it influences the position of the mean fit value of $\omega_b$ and $\omega_{cdm}$, consequently leading to a deviation on $\Omega_m$.

Our model exhibit a level of precision similar to most works, with the exception of Philcox \cite{Philcox_2020} and Chen \cite{Chen_2022}, who report narrower constraints than ours. This is attributed to that both studies incorporate geometrical information from the post-reconstruction of BAO in Fourier and Configuration space, which helps tighten their constraints. Additionally, Tröster \citep{Troster_2020} present slightly broader constraints compared to our results, which we attribute to their use of broader $\omega_b$ priors. 

We conclude that our results with EFT-GSM are in agreement with other full-shape analyses, we found differences within 1-2$\sigma$ ($1.7\sigma$  D'Amico, $1.6\sigma$  Troster and Zhang), this level of agreement can be attributed to the differences in the samples, number of free cosmological parameters, and priors. Therefore, we consider that our EFT-GSM model is a competitive and robust configuration space analysis, that can serve as a complement to other Fourier space methodologies.
 
\subsection{Extensions to Baseline analysis}
\label{complementary_analysis}
We have introduced our EFT-GSM methodology and demonstrated its capability to accurately recover the cosmology of the NSERIES simulations when assuming an error magnitude similar to that expected from future surveys like DESI. Additionally, we applied our methodology to the BOSS data and found that the results we obtained were consistent with those reported by others groups doing full shape analyses with BOSS data. We are now interested in running our methodology using different configurations of our model. This can teach us how various aspects of our methodology impact our final constraints on the parameters.
Our first test involves exploring the capability of our model to constrain cosmological parameters when we modify the priors of two key cosmological parameters $n_s$, and $\omega_{b}$.

It is common practice, when conducting clustering analysis of large-scale structure (LSS), to constrain the values of certain cosmological parameters that are poorly constrained using LSS with external observables. With this in mind, in our baseline analysis, we held $n_s$ constant with a value specified in Table \ref{tab:gsm_priors}, derived from CMB experiments. We also imposed restrictive priors on $\omega_b$. These priors were estimated by measuring the deuterium to hydrogen abundance ratio in a near-pristine absorption system toward a quasar. By assuming a reaction cross-section between deuterium and Helium-3, one can determine strong constraints on $\omega_b$ values. We refer to these priors as Big Bang Nucleosynthesis (BBN) priors throughout this work. 

Here, we explore the constrains we obtain on the cosmological parameters when extending the analysis in these two cosmological parameters, by relaxing the priors on $\omega_{b}$ and  letting $n_s$ free:
 \begin{align}
   \omega_b &: \, \mathcal{N}[0.02237,0.00037] \\
   n_s &: \,  \mathcal{U}[0.5,1.5]   
 \end{align}
 
 The results of these analysis are shown in Table \ref{tabl_wb_ns} and Figure \ref{fig:priors}. We note that, when comparing yellow (BBN prior) posteriors/contours with green ($10 \times$ BBN priors), that in general widening the priors on $\omega_b$ reduces the precision of all other cosmological parameters in particular in $h$ and $\omega_{cdm}$, the error is 2 and 1.6 times larger, although there are no significant shifts of the central values of the posteriors. 
 
This is consistent with the results reported in Tröster \cite{Troster_2020}, they use priors of around 10 times the BBN results and find wider posteriors than other reanalysis of the BOSS DR12 data. This is shown in figure \ref{comparison_b13_papers}.

Ivanov \cite{Ivanov:2019pdj} also investigated the effect of varying the priors on $\omega_b$, finding significantly weaker constraints on $h$ and milder effects on $\Omega_m$, this is consistent with our results as less constraining power in $\omega_{cdm}$ translates to $\omega_m$. 
Brieden \cite{Brieden_2022d} also explored extending the priors in Full Shape (and ShapeFit) analysis. They find that in their Full Shape fits the constraints on $\omega_{b}$ derived from the amplitude of the BAO depends on the ratio $\omega_b/\omega_{cdm}$. Therefore, in the prior-dominated regime the tight constraints on $\omega_b$ helps to fix the shape and narrows the posterior of  $\omega_{cdm}$. When using wider priors the ability of the model to fit the amplitude of the BAO drives the accuracy of the fitting results.

\begin{figure*}
 \begin{center}
\includegraphics[width=90mm]{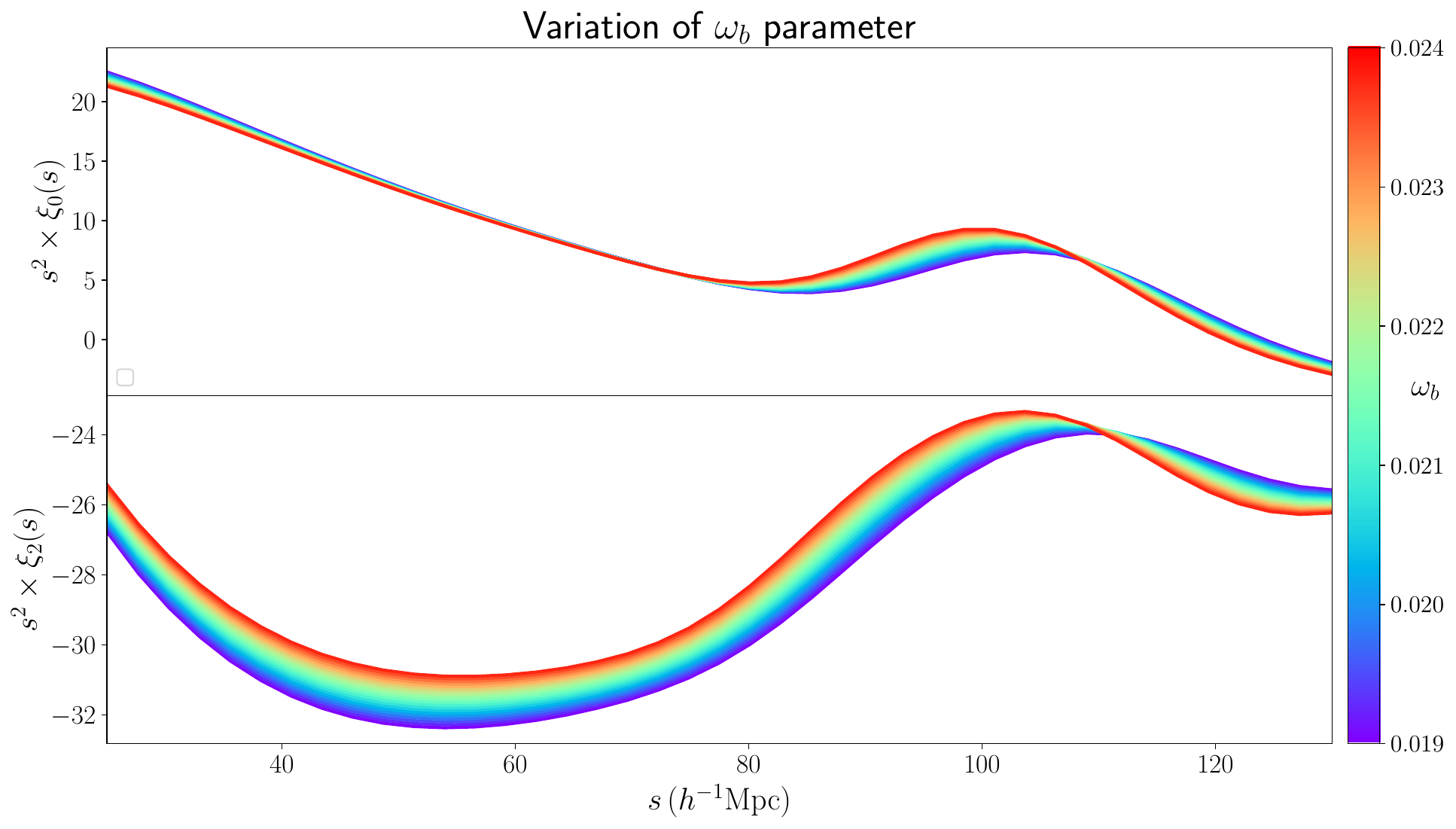}
\caption{ Effect of varying $\omega_b$ in multipoles, top monopole and bottom quadrupole, keeping fix $\omega_m$.  }
\label{fig:omegab_multipoles} 
 \end{center}
\end{figure*}

\begin{figure*}
 	\begin{center}

\includegraphics[width=110mm]{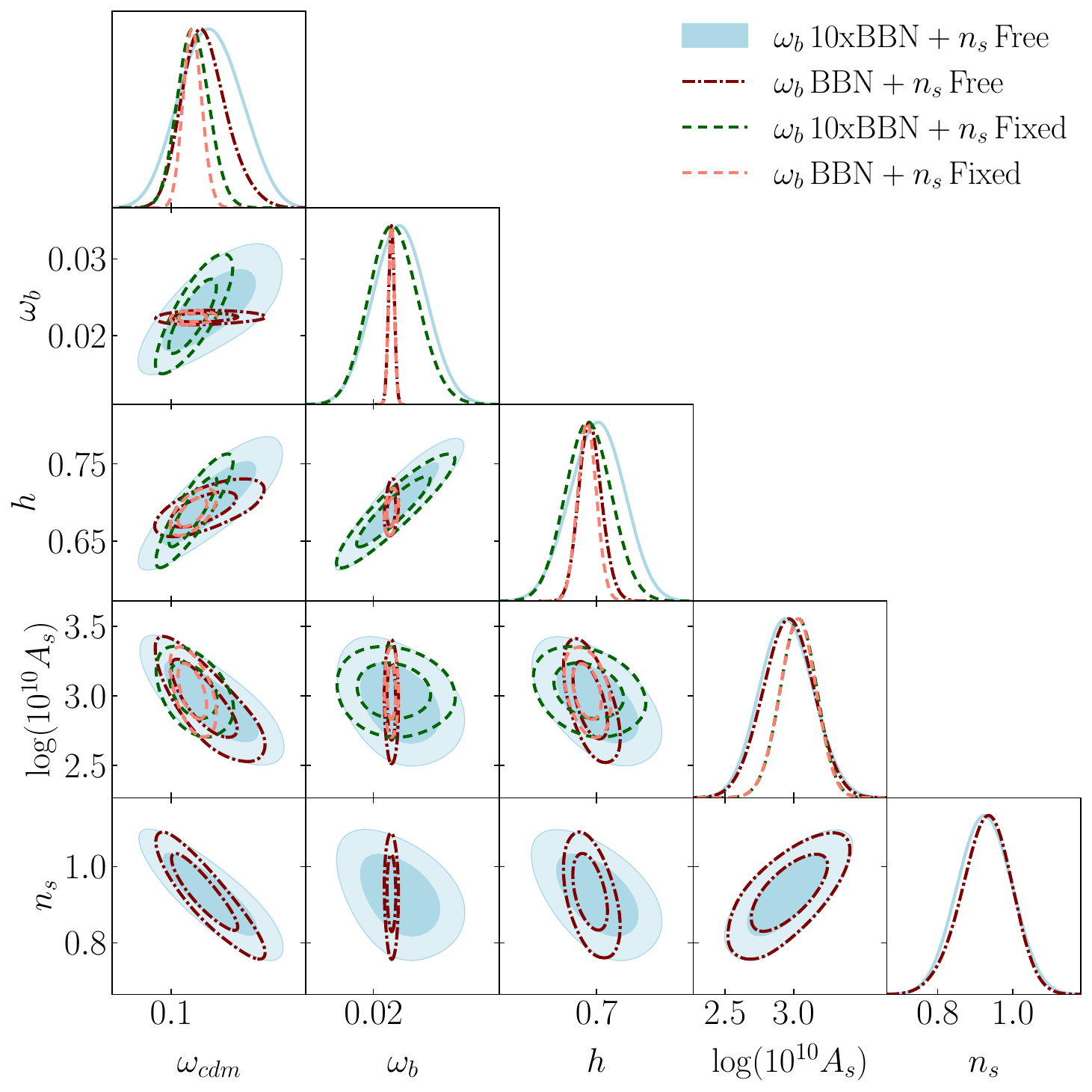}
\caption{
Triangular plot illustrates the likelihood contours obtained when applying our methodology with various configurations of prior settings for the cosmological parameters $\omega_b$ and $n_s$. These fittings were conducted using the BOSS $z_1+z_3$ dataset. The yellow contours represent our standard prior configuration. The green and red contours correspond to wider priors for $\omega_b$ and $n_s$ respectively, while keeping the other parameters at their standard priors. The blue contours correspond to both priors set to wider values. }
\label{fig:priors}
 	\end{center}
\end{figure*}

We would like to highligth that in the case of the configuration space multipoles the effect of varying $\omega_b$ is not isolated in the shape or position of the BAO peak as shown in Figure \ref{fig:omegab_multipoles}.

We also analyze the effect of varying the parameter $n_s$, which as stated above is originally fixed to the Planck value in our baseline analysis. By comparing the yellow ($n_s$ fixed) and magenta ($n_s$ with a flat prior) contours in Figure \ref{fig:priors} we note that fixing $n_s$ has a strong effect on the precision of $\omega_{cdm}$ but a smaller effect on $h$ (as been observed in previous analysis in the Fourier space \citep{Ivanov:2019pdj}), the rational is that $n_s$ and $\omega_{cdm}$ information is coming from the slope, thus again fixing the shape contributes to find tighter constraints on $\omega_{cdm}$. The results are shown in Table \ref{tabl_wb_ns}, the case with varying $n_s$ and keeping the BBN prior on $\omega_b$ shows 2 times larger errors in $\omega_{cdm}$ and $1.25$ times in $h$ also affecting the constraints in $A_s$ by 1.4 factor in the errors. We observe as well that with free $n_s$, the posteriors of $\omega_{cdm}$ and $h$ are shifted towards higher values but still consistent between them within $1$ $\sigma$, this behavior is also consistent with previous analysis in Fourier space where shifts of $1$ $\sigma$ and 0.5$\sigma$ in $\Omega_m$ and $H_0$ respectively \cite{Ivanov:2019pdj}.

\begin{center}
\begin{table*}
\ra{1.7}
\begin{center}

\begin{tabular} { c c c c c}
\hline
Parameter & $\omega_b$--$\text{10}\sigma$, $n_s$ Free & $\omega_b$--BBN, $n_s$  Free   & $\omega_b$--10$\sigma$, $n_s$  Fixed & $\omega_b$--BBN, $n_s$  Fixed   \\
\hline
$\omega_{cdm}$  & $0.121\pm 0.016 $   & $0.118^{+0.010}_{-0.013}  $ &  $0.1119\pm 0.0083  $        & $0.1115\pm 0.0050$\\

$\omega_{b}$   &  $0.0235\pm 0.0034 $ &$0.02237\pm 0.00037 $  & $0.0227\pm 0.0031$                    &  $0.02237\pm 0.00037$\\

$h $            & $0.701\pm 0.035$     &   $0.692^{+0.014}_{-0.016}$ &  $0.689\pm 0.029$                  &$0.688\pm 0.012$ \\

$\ln (10^{10}A_s)$  &  $2.95\pm 0.19$        & $2.96\pm 0.18 $     & $3.03\pm 0.13$                             & $3.03\pm 0.13$ \\

$n_s$  &  $0.925\pm 0.070$        & $0.930^{+0.070}_{-0.063}$     & $0.97$                             & $0.97$ \\
\hline
\end{tabular}
\caption{0.68 c.i. for the cosmological parameters utilizing the monopole and quadrupole measurements of the $z_1+z_3$ samples when we use a BBN-prior on $\omega_b$ and when we relax that prior to have a standard deviation of $\sigma = 10 \times \sigma_\text{BBN}$ (denoted by $10\sigma$) and $n_s$ to be free.}
\label{tabl_wb_ns}
\end{center}
\end{table*}
\end{center}

\subsection{Exploring the Information Content of Multipoles}
\label{mono-quad}

\begin{figure*}
\begin{center}
\includegraphics[width=110mm]{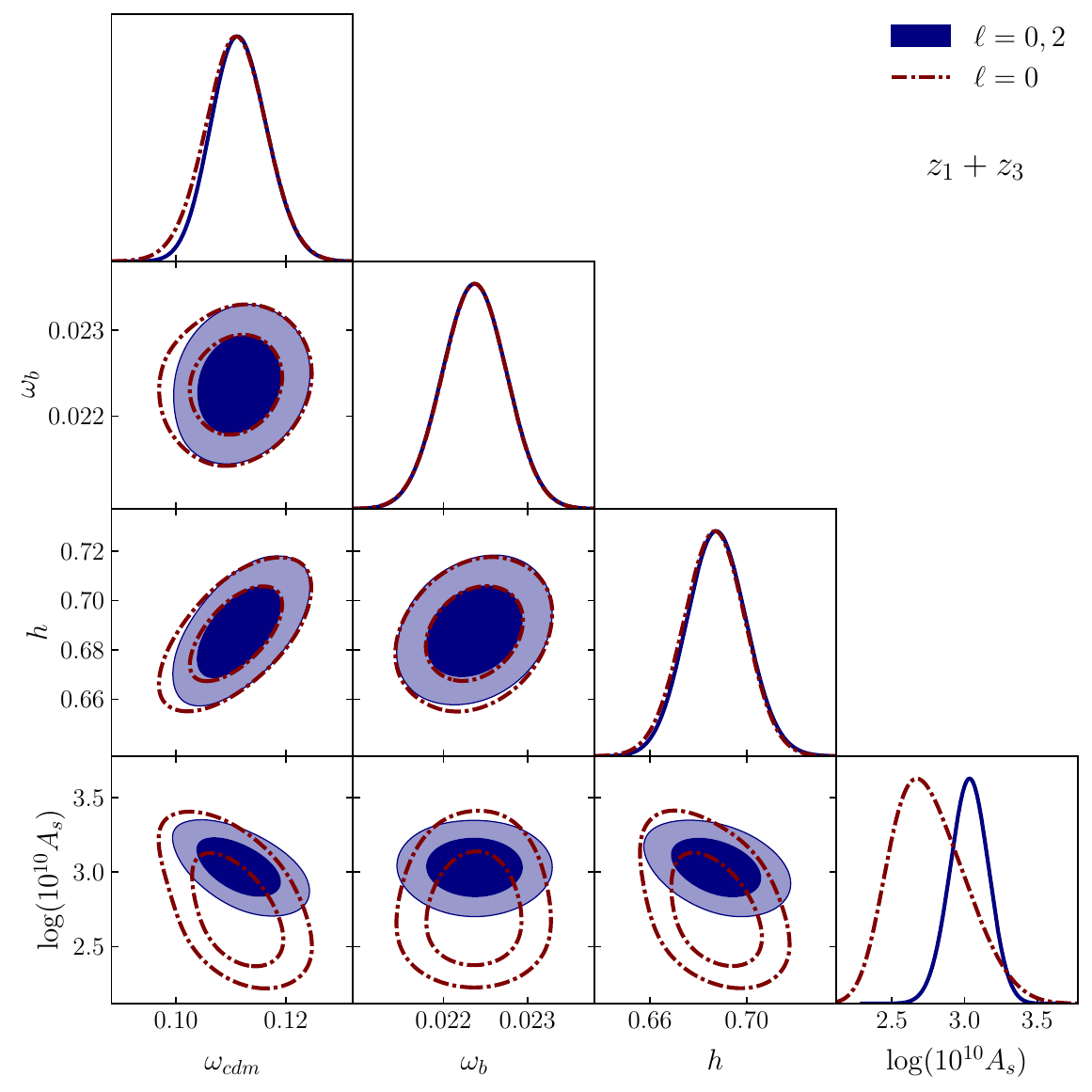}
\caption{Triangular Plot illustrating the likelihood contours obtained using the MCMC methodology when fitting the BOSS dataset $z_1+z_3$ with two different configurations of our model. The blue histograms and contours represent the results of fitting both the monopole and the quadrupole of the correlation function, while the red histograms and contours depict the results of fitting only the monopole.}
\label{mono_vs_monoquad_triangular}
\end{center}
\end{figure*}

This last section we explore the information content and constraining power of the multipoles.
Our last test consists of running a new MCMC fit on the $z_1+z_3$ dataset using our standard configuration. However, this time, we only fit the monopole of the correlation function. Figure \ref{mono_vs_monoquad_triangular} displays the results of this monopole-only fit (red dashed lines) and compare to our baseline analysis, which utilizes both the monopole and quadrupole (blue lines and filled contours).

The results for both cases are summarized in Table \ref{mono}. Interestingly, we observe that the monopole-only approach is capable of recovering our core cosmological parameters, namely $\omega_{cdm}$ and $h$, with nearly the same level of accuracy ($\Delta {\omega_{cdm}}=0.0006$, and $\Delta{h}=0.001$) and precision ($\Delta \sigma_{\omega_{cdm}}=0.0005$, and $\Delta\sigma_{h}<0.001$) as when including the quadrupole. As expected, most of the valuable cosmological information resides within the monopole of the correlation function. However, $A_s$ becomes poorly constrained. 

This is also expected, because RSD, mainly affects the amplitude of the quadrupole to monopole ratio at large scales, which breaks the degeneracy in the parameter $\beta\equiv f/b_1$. Since $A_s$ is highly degenerate with the large-scale bias, the inclusion of the quadrupole induces tighter estimations on $A_s$. The results obtained in this section are expected on theoretical grounds. However, the quadrupole also contains information on the BAO scales, and one would expect that this will translate on better estimation of $\omega_{cdm}$ and $h$, perhaps only a small improvement. Nevertheless, according to our results, the latter is not happening at all, which we find  sligthly surprising. 

\begin{center}
\begin{table*}
\ra{1.7}
\begin{center}

\begin{tabular} {  c c c ccc}

&&$\text{BOSS Combined z1+z3}$ &&\\
\hline
&$\omega_{cdm}$ &$\omega_{b}$ & $h $ &  $ \mathrm{ln}(10^{10}A_s)$ \\
\hline
 $\xi_0$  & $0.1109\pm 0.0055 $   & $0.02237\pm 0.00037  $ &  $0.687\pm 0.012 $        & $2.75^{+0.21}_{-0.29}$ \\
$\xi_0+\xi_2$&$0.1115\pm 0.0050$&$0.02237 \pm  0.00037$ &$0.688\pm 0.012$&$3.03\pm 0.13       $\\
\hline
\end{tabular}
\caption{Summary of fitting   combined samples z1 and z3 from  BOSS using only monopole compared with our baseline analysis with monopole and quadrupole}.\label{mono}

\end{center}
\end{table*}
\end{center}

\section{Conclusions}
There are two distinct philosophies for extracting cosmological information from the shape of the 2PS of LSS. In the first approach, denoted as the compressed methodology, the cosmological template is fixed and fits are done over a small set of compressed variables related to the BAO and RSD observables. By construction,  the compressed methodology is designed to be more agnostic about the model but offers less modeling freedom. 

In the second approach, denoted as full modeling or full shape modeling, the fits are done with a varying template where all the parameters of an \textit{a priori}
chosen model are simultaneously fitted, including the cosmological parameters.  Full Modelling has shown more constraining power compared with  the classical compressed approaches. However, it is naturally more costly in computational time, even if in recent
years, several methods that make full shape analysis efficient have been developed. 
Extensions of the compressed methodology have been proposed as well, achieving similar levels of accuracy than full shape methodology. 
Since these methodologies complement each other and have different strengths and weaknesses, stage IV experiments are currently working to determine the optimal methodology for extracting cosmological information.

In this work we focused on investigating the full shape methodology in configuration space. 
Until now, most of the analyzes of last-generation surveys with a full-shape
methodology has been developed in Fourier space. Therefore, there is an incentive to explore 
 full-shape analysis in configuration space. 
We present a full-shape analysis of the BOSS DR12 galaxy sample two-point correlation function. Our goal was two-folded: 1) to explore the potential of configuration space analysis and contrast it with its fourier space counterpart, and 2) to show the efficiency and robustness of using neural network acceleration for analysing real data.

In order to analyze the anisotropic clustering signal in configuration space we use an EFT-GSM model, to build second-order perturbation theory templates of the correlation function. While the running time of our model implementation is relatively short (on the order of two seconds), executing a complete MCMC chain using our current EFT-GSM model implementation would require approximately 48 hours with 128 CPUs, due to the substantial number of evaluations required. This represents a significant computational expense. To alleviate the computational cost of our
methodology, we employ neural network emulators to construct surrogate models of our EFT-GSM templates. These neural networks are significantly faster to execute and can converge
in as little as 15 minutes when using the same 128 CPUs.

We performed a systematic validation of our methodology in two categories: 
\begin{enumerate}\item \textit{Model Accuracy}. We tested the ability of our methodology to reproduce the cosmological values of the high-resolution NSERIES simulation correspondant to an $V_{\mathrm{eff}}=40$ $h^{-1}\mathrm{Mpc}$. We tested three minimum scales: $s_{\text{min}}=20$, $30$, $40$ $h^{-1}\mathrm{Mpc}$. Our conclusion is that by utilizing a minimum scale of $s_{\text{min}}=20h^{-1}\mathrm{Mpc}$, we maximize the accuracy and precision of our methodology. Additionally, the predicted value of $A_s$ only agrees with the true value to within $1\sigma$ at this scale. The cosmological parameter estimation is the least accurate when $s_{\text{min}}=40h^{-1}\mathrm{Mpc}$, which can be attributed to missing the data bins with the smaller error bars. 

\item \textit{Emulator Accuracy}. The models presented in this work do not directly use the EFT-GSM model. Instead, we employ neural networks to construct surrogate models of the multipoles. We assessed the ability of these surrogate models to reproduce the true predictions made by the full EFT-GSM model. We tested this by constructing a test set of points in parameter space. We calculated the multipoles using both the EFT-GSM code and the surrogate model independently. We observed that the percentage difference between these models is usually less than $1\%$, and for most models, it's closer to $0.1\%$. Furthermore, we noticed that the MCMC fits generated using our surrogate models provide parameter estimations that are virtually indistinguishable from those produced by the full model.
\end{enumerate}

After validating the methodology, we conducted fits to the BOSS data using our baseline analysis. We used the combined sample used in BOSS DR12 final analysis, and we fitted separately the redshift bins $z_1$ and $z_3$ respectively, while the final fit was built on both bins fitted simultaneously. The fit including both bins resulted in slightly tighter constraints on the cosmological parameters. The measured values of the cosmological parameters are in agreement with each other across all three samples, within 1$\sigma$, with the only exception being the predicted value of $A_s$ between $z_1$ and $z_3$, which agrees at the 2$\sigma$ level.
with the combined sample having constrains on $h$ and $A_s$ that are $\sim 25\%$ smaller than on the individual samples, and $\sim 33\%$ smaller that the $\omega_{cdm}$ constrain of the $z_3$ bin constrain. 

We compared our results with previous full-shape analysis performed on BOSS data. We include in the comparison six full-shape methodologies: five of them in Fourier space \citep{DAmico:2019fhj},\citep{Ivanov:2019pdj},  \citep{Philcox_2020},\citep{Troster_2020},\citep{Chen_2022} and one in configuration space \citep{Zhang2022}. We also compare our results with those obtained using the Shape Fit methodology \citep{brieden_2021b}. We find that our predictions for both $H_0$ and $A_s$ agree within $1\sigma$ with the results from all seven works we compare with. Our predictions for $\Omega_m$ agree within $1\sigma$ with four out of the seven works, but we only agree within $2\sigma$ with the remaining three. We propose that these tensions can be explained by two of these three works using broader priors in $n_s$, and by the other work using a slightly different dataset. We also notice that our constraints have a level of precision comparable to that of five out of the seven works. The remaining two works included post-reconstruction information of the power spectrum and are therefore able to achieve better precision than us.

We  performed 
complementary tests to gain a better understanding of the impact of priors on our constraints. 
We have explored extending the baseline analysis by
relaxing the priors on $\omega_b$ by 10 times the current range $\mathcal{N}[0.02237,0.00037]$ and by letting $n_s$ be a free
parameter with a flat prior of [0.5, 1.5]. 
When we relax the priors on $\omega_b$, we find significantly weaker constraints on $h$ and, a milder effect on $\omega_{cdm}$. When we vary $n_s$ we note a strong effect on the precision of $\omega_{cdm}$ but a smaller effect on $h$, this is due to $n_s$ and $\omega_{cdm}$ having a strong effect on the slope of the multipoles. All of these observations are consistent with what other works have found.

Finally, we explored the information content of the multipoles. We conducted our standard fit using only the monopole of the correlation function and compared it with our baseline analysis that includes both the monopole and quadrupole. We discovered that the monopole-only fit already provides constraints on $\omega_{cdm}$ and $h$ with similar accuracy and precision as when using both multipoles. This suggests that the majority of relevant cosmological information is contained in the monopole of the correlation function, which we find slightly surprising, given that the quadrupole also contains some BAO information. We also noted that the constraints on $A_s$ do worsen significantly, as expected.

\acknowledgments
This work was supported by the high-performance computing
clusters Seondeok at the Korea Astronomy and Space Science Institute. MV, SR and SF acknowledges PAPIIT IN108321, PAPIITA103421, PAPIIT116024 and PAPIIT-IN115424. MV acknowledges CONACyT grant A1-S-1351. This research was partially sup-
ported through computational and human resources provided by the LAMOD UNAM project
through the clusters Atocatl and Tochtli. LAMOD is a collaborative effort between the IA,
ICN and IQ institutes at UNAM.  AA is supported by Ciencia de Frontera grant No.~319359, and also acknowledges partial support to grants Ciencia de Frontera 102958 and CONACyT 283151.

\bibliographystyle{JHEP}
\bibliography{NN_unam}
\end{document}